\documentclass[reprint,amsmath,amssymb,aps,twocolumn,floatfix]{revtex4-2}
\usepackage{graphicx}
\usepackage{dcolumn}%
\usepackage{bm}
\usepackage{relsize}
\usepackage{mathrsfs}
\usepackage{multirow}
\usepackage{textcomp}
\usepackage{textcase}
\usepackage{siunitx}
\usepackage{amsmath}
\usepackage{hyperref}
\usepackage{amssymb}
\usepackage{footnote}
\usepackage{floatrow}
\usepackage{caption, threeparttable}
\usepackage{threeparttable}
\usepackage[utf8]{inputenc}
\usepackage[dvipsnames]{xcolor}
\usepackage[utf8]{inputenc}

\newcommand{\dd}{\mathnormal{d}}
\newcommand{\old}[1]{}
\DeclareUnicodeCharacter{2212}{-}

\begin{document}
\preprint{APS/123-QED}
\title{Structure of the thin disk in the background of a distorted naked singularity versus a distorted static black hole}

\author{Shokoufe Faraji}
\email{shokoufe.faraji@zarm.uni-bremen.de}

\affiliation{%
University of Bremen, Center of Applied Space Technology and Microgravity (ZARM), 28359 Germany}%

\begin{abstract}
We consider the standard relativistic geometrically thin and optically thick accretion disc around a distorted static black hole and a distorted naked singularity. The distortion is the result of the existence of a static axisymmetric external distribution of matter via exercising quadrupole moments. Our main purpose of this work is to investigate whether the naked singularity models have their own observational fingerprint if they do exist. In fact, understanding the astrophysical behavior of naked singularities seems to be extremely important in the perspective of general relativity and understanding of black hole physics today. 
\end{abstract}

\maketitle

\section{Introduction}


Studying accretion disc models is crucial for understanding the emissions observed from astrophysical objects. For example observation of the luminous AGN discs, X-ray binaries, star-forming discs and cataclysmic variables in quiescence.


There are impressive researches on this phenomena, analytically or by using simulation in an complementary fashion. The standard thin disc model was introduced by Shakura \& Sunyaev \cite{1973A&A....24..337S}, Bardeen, Press \& Teukolsky \cite{1972ApJ...178..347B}, Novikov \& Thorne \cite{1973blho.conf..343N} in 1973, and Lynden-Bell \& Pringle 1974 \cite{1974MNRAS.168..603L}. Later, {Paczy{\'n}ski} \& Bisnovatyi-Kogan \cite{1981AcA....31..283P}, Muchotrzeb \& {Paczy{\'n}ski} \cite{1982AcA....32....1M}, and Abramowicz et al \cite{1988ApJ...332..646A}, introduced the slim disc by considering some terms that are neglected in the thin disc models. There are a number of works on the standard accretion discs on the Kerr space-time, (for a review see e.g. \cite{Abramowicz2013}), as well as on the other space-times for example, on the Johannsen-Psaltis parametrised space-time \cite{Chen2012}, on the Kerr-de Sitter space-times \cite{Stuchlik2004}, Boson stars \cite{Torres2002}, wormhole \cite{Harko2009}, quark stars \cite{Kovacs2009}, gravastars \cite{Harko2009c}, and Bose-Einstein condensate stars \cite{Danila2015}. Also, in analogue gravity model \cite{Das_2007} $f(R)$ theories \cite{Pun2008,Perez2013}, Ho{\v{r}}ava gravity \cite{Harko2011}, Chern-Simons modified gravity \cite{Harko2010}, scalar-vector-tensor theory \cite{Perez2017}, an higher dimensional Kaluza-Klein black hole \cite{Chen2012}, and Gibbons-Maeda-Garfinkle-Horowitz-Strominger charged black holes \cite{Karimov2018}.

In astrophysical study, most of the models for the central body, both within general relativity and alternative theories of gravity, assume that the space-time outside the object does not contain any additional matter. While in a real astrophysical environment, a black hole due to its strong gravitational field, is not necessarily isolated but in particular surrounded by the additional matter, radiation and electromagnetic fields. However, these models require to have asymptotic flatness space-time, which is equivalent to the restriction to the cases that are isolated in the space. Even though it is fundamental that restrict our attention, in the first instance, to isolated compact objects, the question as to how they maybe distorted by external distribution of mass also can be serving as the basis for a quest. However, the investigation of this type of systems, either by analytical or numerical tools, may fall back on the ability to set up suitable representation based on physical assumptions. In this regard, one may consider the external matter distribution as the contribution of the outer part of the disc, where self-gravity seems to manifest. Indeed, the gravitational field produced by the disc itself may have an observational effect in studying the active galactic nuclei \cite{Lodato2007}. There are a variety of studies on black hole-disc systems e.g. \cite{Will1974,PhysRevD.49.5135}.

Considering an external gravitational field up to a quadrupole add to the Schwarzschild space-time was introduced in 1969 by Doroshkevich and the colleagues. Later in 1982, a detailed analysis of the global properties of the distorted Schwarzschild space-time was published in 1982, by Geroch \& Hartle \cite{1982JMP....23..680G}. In 1997, the explicit form of the metric was presented in \cite{1997PhLA..230....7B}. During the last years a lot of studies have been devoted to multipole moment of external matter in this sense, for example distorted black holes in terms of multipole moments \cite{PhysRevD.34.3633}, arbitrarily deformed Kerr-Newman black hole in an external gravitational field \cite{PhysRevD.57.3382}, distortion of magnetized black hole \cite{PhysRevD.96.024017}, among many others. We explain more about this solution in Section \ref{sec:Dnaked}.


In this work, we investigated the properties of the thin model of the accretion disc via exercising quadrupole moments in an effort to investigate how much the thin discs around a naked singularity deviate from the accretion disc around a static black hole in the presence of static and axially symmetric external distribution of matter.

Beyond the shadow of a doubt, one of the most fundamental issues in relativistic astrophysics is the final stage of a massive object. In the scope of general relativity, it is usually assumed that the answer is formation of a regular black hole, but the possibility of the formation of a naked singularity is also not ruled out. Although the EHT images may eliminate some black hole alternatives, still there are other solutions in general relativity without an event horizon that can imitate the electromagnetic characteristic of a black hole. Such observational data would surely provoke a challenging identification of the primary object \cite{2002A&A...396L..31A,PhysRevD.78.024040,2019MNRAS.482...52S}. In this respect, this work is one step in a series of efforts toward comparing the astrophysical properties of a naked singularity with a regular black hole.

Even if naked singularities form hypothetically in nature, it is important to characterize them through their observational signatures from black holes. In this respect, naked singularities have been considered in many studies in such a way that most of them are resulting from gravitational collapse. We briefly discuss naked singularity and some of the related work in Section \ref{sec:naked}. In this paper, we consider the distorted naked singularity which is as a result of two proper quadrupole moments. In fact, this metric is the simplest generalization of the distorted Schwarzschild solution and $\rm q$-metric, containing two quadrupole parameters and may consider as the local $\rm q$-metric \cite{2020arXiv201015723F}. In this way, naked singularities may have a clear interpretation in terms of their quadrupole moments.

This procedure, may open the opportunity of thinking of quadrupoles as additional physical degrees of freedom that facilitates the search for connecting observational phenomena like the gravitational wave generated by an non-isolated, self-gravitating axisymmetric distribution of mass, to the properties of the central object and its exterior. In particular, it seems that the present understanding of the black hole accretion phenomenon mostly relies on the studies of stationary and axially symmetric models. Consequently, the geometric configuration of an accretion disk located around this space-time depends explicitly on the value of the both quadrupole parameters in such a way that it is always possible to distinguish between a distorted static black hole and a naked singularity.


This paper is organized as follows. In Section \ref{sec:naked}, there is a brief overview on the naked singularities. Then, in Section \ref{sec:Dnaked} we explain the $\rm q$-metric and the space-time of the distorted naked singularity in the presence of an external static and axially symmetric matter. Then we revisit and discuss in Section \ref{sec:disc} the relativistic thin disc model. In Section \ref{sec:eq} the construction of the thin disc around the distorted naked singularity is explained. The solution and results are presented in Section \ref{sec:results}. Finally, summary and conclusions are in Section \ref{sec:discuss}.

Throughout this paper, the geometrized units, $c=1$ and $G=1$, have been used. If somewhere other units are used it is specified there.







\section{Naked singularity}\label{sec:naked}
In general relativity, a singularity is a region in the space-time where the curvature becomes infinite and the space-time manifold is not smooth. Mathematically, the singularity is the boundary of space-time, and not regarded as a point in the space-time manifold. 

By definition, a black hole is a singularity that is covered by at least one horizon. According to the Penrose-Hawking singularity theorem, a singularity is unavoidable under some general conditions \cite{PhysRevLett.14.57,hawking_ellis_1973}. But this theorem does not say any detail about features of the singularities. In fact, if the singularity is not covered by a horizon then it may be observable for an observer at infinity. This kind of singularity, in general, is called naked singularity that leads to several strange properties of the space-time mainly regarding stability and causality \footnote{Moreover, the concept of entropy for a black hole is defining by the means of the area of the horizon. However, in the case of naked singularities, it is not possible to define and would violate Bekenstein's generalized second law \cite{PhysRevD.9.3292}.}.

In the gravitational collapse, if a pressure gradient force on a body is not sufficiently strong, a body can continue collapsing due to the self-gravity. The final outcome of such a collapse is either a black hole or a naked singularity, depending on the initial density configuration and velocity distribution of the particles.

There are a plenty of researches over the years that admitted the appearance of such a singularity at the final stage, for example, the seminal works on Einstein cluster geometry \cite{KumarDatta1970,Bondi1971}, on the spherical collapse of a perfect
fluid, numerically e.g. \cite{PhysRevD.42.1068,Harada_1998} and analytically e.g. \cite{PhysRevD.45.2147}. On the Vaidya null dust solutions \cite{Dwivedi_1991}, Tolman-Bondi dust solutions \cite{Magli_1997,Christodoulou1984} among many more.

For the homogeneous dust collapse case, the trapped surfaces \footnote{A trapped surface is a two-dimensional surface such that both the congruence of future directed null geodesics normal to the surface are converging.} and apparent horizon are formed before the formation of the singularity, and therefore the singularity is covered. On the other hand, for inhomogeneous collapse, there are classes of initial conditions for which the singularity forms before the formation of the trapped surfaces and apparent horizon. Therefore, in this case it will end up with a naked singularity e.g. \cite{Singh_1996,PhysRevD.62.044001,PhysRevD.47.5357,2019MNRAS.482...52S,Joshi_2013,2011IJMPD..20.2641J}. 

Most of the studies on naked singularity formation in this area, were assumed spherical symmetric configurations of matter, first for simplicity and reduce the dimensionality of the problem, second a spherically symmetric distribution of matter cannot emit gravitational radiation, like a spherically symmetric distribution of electric charges that cannot emit electromagnetic radiation. In the absence of this symmetry, any moving mass radiate gravity waves, which complicates the dynamical equations.

In addition, in the study of naked singularities, there is a family of the solution with a proper quadrupole moment that makes the horizon singular, including \cite{osti_4201189,2011IJMPD..20.1779Q,Toktarbay_2014}.


In 1969 Penrose asked a question: "Does there exist a cosmic censor who forbids the appearance of naked singularities, clothing each one in an absolute event horizon?" This leads to the weak cosmic censorship conjecture  \cite{Penrose2002,1986CaJPh..64..120I},  which requires that the singularity must not be naked and visible from the future null infinity. In another word, the maximal Cauchy development possesses a complete future null infinity. There is also another version, namely strong cosmic censorship conjecture whose demands the singularity must be spacelike.

Many studies have been so far devoted to verifying this conjecture, for example by numerical simulations \cite{PhysRevLett.66.994,10.2307/29774425} or by gedanken experiments in such a way that the possibility of turning a black hole to a naked singularity was been verified by plunging a test particle with a large electric charge and arbitrary angular momentum into an extreme black hole \cite{1974AnPhy..82..548W,Semiz_1990,PhysRevD.22.791,1999PhRvD..59f4013H,Chirco_2010,Sorce_2017,Richartz_2008,Matsas_2007,Isoyama_2011}. Also, there are some work including the back reaction effects. However in this case, the outcome result for this conjecture depends on the interpretation of the result \cite{Hod_2008,Matsas_2009}. Also, including the self-force of a test body in \cite{Barausse_2010}. In \cite{Joshi_2002}, the authors claim that if shearing effects is sufficiently strong near the singularity, delay the formation of the apparent horizon so the singularity becomes visible, however, if the shear is weak we end up with a black hole.

In spite of a huge amount of effort, this remains still a conjuncture. While, regarding the existence of counterexamples, it could not be proved in generality. However, it may be possible that the weak cosmic censorship conjecture would be preserved, if the naked singularity will evolve into a compact object in a finite time \cite{Penrose2002,hioki2019dynamical}. Besides, Regarding the scale where the singularity is forming, the classical relativity methods in this concept may need to be modified by the ongoing quantum theory of gravity e.g. \cite{Penrose2002}.

Also, there is a conjuncture by Thorne, called hoop conjectured \cite{1972mwm..book..231T,1973grav.book.....M} that a horizon forms if and only if a collapsing mass is sufficiently compacted in all dimensions. Thus, no one dimension of the collapsing object can be notably larger than others.

Interestingly, in the frame of loop quantum gravity also, a naked singularity could exist; however, its physical meaning is slightly different e.g. \cite{Bojowald2008}. Also, the interesting approach in the black hole information problem implies that the firewall is needed to break the surplus entanglements among the Hawking photons. Classically, the firewall becomes a naked singularity \cite{2017NatSR...7.4000Z}. Besides, there has been a series of discussions on possible violation of cosmic censorship in asymptotically AdS space-times e.g \cite{2017PhRvL.118r1101C}.


In the view of the visibility of naked singularity to the far away observers, it is the possibility of having observable signatures from ultra-strong gravity near the singularity. 
In fact, the black holes and naked singularity models would have distinguishable physical properties.


In recent times, much attention has been devoted to the observational properties of black holes and naked singularities that make them distinguishable, either vacuum solutions with naked singularities or collapse solutions in the presence of matter, including through iron line emission \cite{Liu_2018,Bambi_2013}, and through accretion disc properties like luminosity and study its spectrum \cite{Joshi_2013}. In later work, they constructed a toy model of static spherically symmetric perfect fluid interiors with a singularity at the origin, in which it is matched to a Schwarzschild exterior geometry. In this matter model, the radiant energy flux and spectral energy distribution are much greater for naked singularity than the same size black hole. In \cite{PhysRevD.82.124047} accretion disc around Kerr-type black hole and rotation naked singularity was studied. They found out the major difference between rotating naked singularities and Kerr black holes is in the frame-dragging effect. In \cite{2019MNRAS.482...52S}, features of naked singularities studied via comparing shadows of the Schwarzschild black holes with the shadows of two classes of naked singularities that result from the gravitational collapse of spherically symmetric matter, also by lensing \cite{Gyulchev_2008}, and studying timelike geodesics \cite{PhysRevD.93.024024,Bambhaniya_2019}. Study of the bound orbit near a scalar field black hole and scalar field singularities in \cite{2019EPJC...79..709P}. It has been shown that the shape and parameters of a bound orbit depend crucially on the type of configuration. In \cite{Iguchi_1999,Iguchi_2000}, the gravitational radiation from a naked singularity is studied. Equilibrium configuration from gravitational collapse is studied in \cite{Joshi_2011}, they showed that the gravitational collapse of a matter cloud with non-vanishing tangential pressure, from regular initial data, can give rise to a variety of equilibrium configurations as the collapse final state. Also, Investigation of Hawking radiation emitted from a naked singularity has been studied in \cite{2017NatSR...7.4000Z}.  

From such a perspective, in the present work, we examined the properties of accretion disc around a naked singularity that results from adding a relatively small quadrupole moment to the Schwarzschild solution, in the presence of the external distribution of matter. Indeed, there are significant differences compared to discs around a static black hole that can help to distinguish them in nature if they exist. In the next Section we briefly describe this space-time which was discussed in \cite{2020arXiv201015723F}.

\section{Distorted naked singularity}\label{sec:Dnaked}
The general static axisymmetric solution of the vacuum Einstein equations in four and five dimensions is given by the Weyl metric. For a higher dimension, this solution is given by generalized Weyl metric \cite{Emparan_2002}. The Weyl metric reads as

\begin{align}\label{weylmetric}
\dd s^2 & =-e^{2\psi}\dd t^2 +e^{2(\gamma-\psi)}(\dd\rho^2+\dd z^2) +e^{-2\psi}\rho^2 \dd\phi^2\,,
\end{align}
where metric functions $\psi$ and $\gamma$, only depend on $\rho$ and $z$.
The function $\psi$ plays the role of a gravitational potential that in the flat space

\begin{align}
ds^2=d\rho^2+dz^2+\rho^2 d\phi^2,
\end{align}
obeys the Laplace equation,

\begin{align}
\psi_{s_,\rho\rho}+\frac{1}{\rho}\psi_{s_,\rho}+\psi_{s_,zz}=0.\label{laplap}
\end{align}
The function $\gamma$ is simply obtained by an integration of the explicit form of function $\psi$ \footnote{The Laplace equation for $\psi$, is the integrability condition for $\gamma$.}, 

\begin{align}\label{gammaeq}
\gamma_{,\rho}=&\rho(\psi^2_{,\rho}-\psi^2_{,z}),\nonumber\\
\gamma_{,z}=&2\rho \psi_{,\rho}\psi_{,z}.\
\end{align}
The relation between Schwarzschild coordinates and Weyl coordinates is given by

\begin{align}
\rho&=\sqrt{r(r-2M)}\sin{\theta},\\
z&=(r-M)\cos{\theta},\
\end{align}
where $M$, is a parameter which can be identified as the mass of the body generating the field, which is expressed in dimension of length. In 1959 Erez and Rosen have pointed out that the multipole structure of a static axially symmetric solution has a simpler form in elliptical coordinates rather than the cylindrical coordinates of Weyl \cite{osti_4201189,PhysRevD.39.2904}. 
Then by rewriting Weyl metric \eqref{weylmetric} in the prolate spheroidal coordinates $(t, x, y, \phi)$ \footnote{Prolate spheroidal coordinates are a three-dimensional orthogonal coordinates that result from rotating the two-dimensional elliptic coordinates about the focal axis of the ellipse.} we obtain,


    \begin{align}\label{weylxy}
	\dd s^2 =& - e^{2\psi} \dd t^2+ M^2e^{-2\psi}\\
	&\left[ e^{2\gamma} \left( \frac{\dd x^2}{x^2-1}+ \frac{\dd y^2}{1-y^2} \right) + (x^2-1)(1-y^2)\dd{\phi}^2\right],\nonumber\
\end{align}
where $t \in (-\infty, +\infty)$, $x \in (1, +\infty)$, $y \in [-1,1]$, and $\phi \in [0, 2\pi)$.
The relation of this coordinates to the Weyl coordinates are as follows,
\begin{align}
    {x} & =\frac{1}{2{M}}(\sqrt{\rho^2+({z}+{M})^2}+\sqrt{\rho^2+({z}-{M})^2})\label{trafo1}\,,\\
    {y} & =\frac{1}{2{M}}(\sqrt{\rho^2+({z}+{M})^2}-\sqrt{\rho^2+({z}-{M})^2})\label{trafo2},\
\end{align}
Equivalently, $\rho={M}\sqrt{({x}^2-1)(1-{y}^2)},$ and ${z}={M}{x}{y}$.
Also, the relation between the prolate spheroidal coordinates $(t, x, y, \phi)$, and the Schwarzschild coordinates $(t, r, \theta, \phi)$ is given by
\begin{align}\label{transf1}
 x =\frac{r}{M}-1 \,, \quad  y= \cos\theta.\,
\end{align}
In what follows we discussed the presence of the external source via quadrupoles. 

In general, there are two classes of solution for the vacuum, static Einstein's equation with a smooth event horizon. The asymptotically flat solution, known as the Schwarzschild space-time, and another class that is not asymptotically flat and known as the local black hole or distorted black hole. This solution is obtained by assuming the black hole is no longer isolated, via considering the existence of a static and axially symmetric external distribution of matter outside the horizon. Then, by its definition, this solution is only valid locally \cite{1982JMP....23..680G,Chandrasekhar:579245}. This approach was introduced first by Geroch and Hartle in 1982 \cite{1982JMP....23..680G}. For its construction the Weyl metric \eqref{weylmetric} is used. As it was mentioned before, the metric function $\psi$ is the solution of the Laplace equation \eqref{laplap}, namely $\psi$ is a harmonic function, also equation \eqref{laplap} is linear in $\psi$. These properties are the key in this method since this allows us to add another harmonic function $\hat{\psi}$ to $\psi$,
 \begin{align}
    \psi = \psi + \hat{\psi}. 
\end{align}
This new term $\hat{\psi}$ is considered as a distortion function or the background field in the paper by Geroch \cite{1982JMP....23..680G}. 
Although, the equation for $\gamma$ is not linear, but it is useful to present $\gamma$ also as 
\begin{align}
   \gamma  = \gamma + \hat{\gamma}.
\end{align}
These distortion functions are obtained in terms of the Legendre polynomial e.g. \cite{BRETON19977,PhysRevD.39.2904}, such that up to quadrupole are given by

\begin{align}
\hat{\psi} & = -\frac{\rm q}{2}\left(-3x^2y^2+x^2+y^2-1\right),\label{psigeneralequ}\\
\hat{\gamma} & = -2x{\rm q}(1-y^2)\nonumber\\
&+\frac{{\rm q}^2}{4}(x^2-1)(1-y^2)(-9x^2y^2+x^2+y^2-1)\label{gammageneralequ},\
\end{align}
where $\rm q$ is the distortion parameter, namely quadrupole for external distribution of matter. Of course, in the case of $\hat{\psi}=0$ and $\hat{\gamma}=0$, we recover Schwarzschild metric \footnote{To have some view of external quadrupole $\rm q$, consider the quadrupole in Newtonian theory. By a convention one can consider a positive external quadrupole as being due to a ring in the equatorial plane, whereas $\rm q<0$ due to this along the z-axis. So, in this picture $\rm q > 0$ would cause a net force radially outward, and $\rm q<0$ cause one radially inward.}.

Aside from this approach of considering extra functions $\hat{\psi}$ and $\hat{\gamma}$ in terms of multipoles, there is another way for applying quadrupole moment within Weyl metric. In this fashion, this static solution with arbitrary quadrupole moment is described in the seminal papers by Weyl \cite{doi:10.1002/andp.19173591804} and Erez and Rosen 1959 \cite{osti_4201189}. Later Zipoy \cite{doi:10.1063/1.1705005} and Voorhees \cite{PhysRevD.2.2119} found a simple solution to consider quadrupole correction to the Schwarzschild metric that can be treated analytically and the resulting metric is known as $\gamma$-metric or $\sigma$-metric, and later via applying a small transformation as $\rm q$-metric \cite{2011IJMPD..20.1779Q}. In 1970 Geroch introduced the relativistic multipole moments of vacuum static asymptotically flat space-time \cite{1970JMP....11.2580G}, and in 1974, it was generalized to the stationary case by Hansen \cite{doi:10.1063/1.1666501}. This area has been discussed extensively in the literature and generalized in many respects. For example, general static axisymmetric solution in prolate spheroidal coordinate \cite{PhysRevD.39.2904}, external field of static deformed mass in \cite{Manko_1990}, explicit multipole moment of static, axisymmetric space-time \cite{Herberthson_2004}. Also, on the stationary case \cite{Toktarbay_2014,doi:10.1002/prop.2190381002,QUEVEDO198513,Castejon_Amenedo_1990}, among many others. Although there are different definitions and approaches to studying multipole moments, they are all physically equivalent \cite{PhysRevD.39.2904}.

The $\rm q$-metric has the central curvature singularity at $x=-1$. Moreover, this space-time is characterized by the presence of a naked singularity situated at a finite distance from the origin at $x=1$, for any chosen value of quadrupole $\rm p$ \cite{doi:10.1063/1.1705005,PhysRevD.2.2119,PhysRevD.93.024024,2011IJMPD..20.1779Q}. However, considering relatively small quadrupole moment, this is located very close to the origin. Of course in the absence of the quadrupole, also in this case the Schwarzschild metric is recovered.
By Geroch definition \cite{1970JMP....11.2580G} for multipole moment and for avoiding a negative mass distribution, the quadrupole is restricted at most to this domain $\rm p\in(-1,\infty)$ \footnote{The lowest independent multipole moments, monopole, for this metric is calculated as $m_{0}=M(1+\rm p)$, that is equivalent to restricting quadrupole at most to this domain $\rm p\in(-1,\infty)$. In fact, the Arnowitt-Deser-Misner mass which characterizes the physical properties of the exact solution also has the same expression and should be positive \cite[App.]{PhysRevD.93.024024}, which for the stationary space-time it is equivalent to Komar mass. Besides, for $\rm p=-1$ one obtains $m_0=0$, meaning all multipole moments vanish and the space-time is flat \cite{PhysRevD.93.024024}.}. It turns out that the only independent parameters are $M$ and $\rm p$ which determine the mass and quadrupole moment \cite{Quevedo:2010vx}. Besides, all odd multipole moments are vanished because of reflection symmetry with respect to the equatorial plane. It is worth mentioning that the difference in the classical and relativistic multipole moments appears at first in the octupole moment. However, we can always choose an origin in such a way that dipole vanishes, then the next term that shows deviation from classical one will be revealed in the 16-pole moment.

Now in this work we consider local $\rm q$-metric, which considers these two approaches together. Ultimately, the metric has this form \cite{2020arXiv201015723F},

\begin{align}\label{EImetric}
	{\rm d}s^2 &= - \left( \frac{x-1}{x+1} \right)^{(1+{\rm p})} e^{2\hat{\psi}} \dd t^2+ M^2(x^2-1) e^{-2\hat{\psi}} \nonumber\\
	 &\left( \frac{x+1}{x-1} \right)^{(1+{\rm p})}\left[ \left(\frac{x^2-1}{x^2-y^2}\right)^{{\rm p}(2+{\rm p})}e^{2\hat{\gamma}}\right.\\
	 &\left. \left( \frac{\dd x^2}{x^2-1}+\frac{\dd y^2}{1-y^2} \right)+(1-y^2) \dd{\phi}^2\right],\
\end{align}
This metric contains three free parameters, namely the total mass, and quadrupole moments $\rm p$, and $\rm q$, which are taken to be relatively small and connected to the $\rm q$-metric and the presence of external mass distribution, respectively. In the case of $\rm q=0$ this turns to the mentioned $\rm q$-metric, and in the case of $\rm p=q=0$ Schwarzschild metric is recovered.

In what follows we review the construction of the thin accretion disc.

\section{Thin accretion disc }\label{sec:disc}

Accretion discs establish when gaseous matter spirals onto a central object by gradually losing its initial angular momentum and form a disc-like configuration in which angular momentum is transported outwards as a result of the differentially rotating fluid, which causes gas to be accreted onto the central object. 	


The standard thin disc is a fascinating analytical model of accretion, was proposed and developed in the seminal works of Shakura \& Sunyaev 1973 \cite{1973A&A....24..337S}, Novikov \& Thorne 1973 \cite{1973blho.conf..343N}, Lynden-Bell \& Pringle 1974 \cite{1974MNRAS.168..603L}. This model has been used to explain a variety of observations where the gas is cold and neutral in such a way that coupling between the magnetic field and the gas is negligible e.g. \cite{1994ApJ...421..163B,mukhopadhyay2005hydrodynamic}. In general, observations provide the luminosity and the maximum temperature of the disc that enable us to fit the model to the data.

In the thin standard accretion model, it is supposed that the disc can radiate a considerable fraction of its rest mass energy locally. This radiation is thermal black body-like radiation which is generated through the viscosity mechanism that we explain later. As a result, the thin disc model is considered as the cold accretion disc. Since, depending on the mass of the central object, the gas temperature is of the order of 100 K that is considered cold regarding the virial temperature. For example, this model cannot produce a very high temperature ($T>10^{10}$ K) observed in the Galactic Center source Sgr $A^{\ast}$ \cite{2007MNRAS.381.1267K}.

One of the crucial assumptions of the standard thin disc is that it is taken to be razor thin and confined to the equatorial plane, meaning the ratio of the disc half-thickness $H = H(r)$ over the radius $r$ is very small, $H/r\ll1$. As a result, the generated heat and radiation losses are in balanced $\rm Q^{gen} = Q^{rad}$, and caused to have a negligible advection, since

\begin{align}
{\rm Q}^{\rm adv}\sim\left(\frac{H}{r}\right)^2{\rm Q}^{\rm gen}.   
\end{align} 
Consequently, it causes to luminosities be approximately below $30\%$ of the Eddington luminosity,

\begin{align}
L_{\rm Edd}:= 1.26\times 10^{38}\left(\frac{M}{M_{\odot}}\right) {\rm \frac{erg}{s}}    
\end{align}
or the mass accretion rate $\dot{M}$ be below the Eddington rate $\dot{M}_{\rm Edd}$ \footnote{There are other definitions of $\dot{M}_{\rm Edd}$ are used in the literature that one can see for example in \cite{Yuan_2014}}. Above this limit, the gas becomes optically too thick and can not radiate all the dissipated energy locally \cite{2006ApJ...652..518M,2008ApJ...676..549S}. 
Therefore, one should justify applying the standard thin model to discs with higher luminosity. Interestingly, also magnetized thin accretion discs in X-ray binaries, at luminosities below 30\% of Eddington, can be described by the thin disc model successfully \cite{2010MNRAS.408..752P}.

The solution to the thin disc model can only be found by applying certain well-known assumptions that the model is based on. Following the approach of Shakura \& Sunyaev \cite{1973A&A....24..337S}, one assumes that the specific internal energy density is negligible and the disc lies in the equatorial plane, implying ${u}^{\theta}$ component of fluid four-velocity vanishes, also quasi-Keplerian circular orbits are assumed with a small radial drift velocity $u^r$, which is much smaller than the angular velocity. Near the central compact object $u^r$ is negative and gives rise to the mass accretion. Also, no mass or angular momentum crosses the disc surfaces. Besides, self-irradiation of the disc, the loss of angular momentum due to wind and radiation are neglected. We consider the sub-Eddington accretion rate which is the proper choice for the thin disc model as it was discussed. 
The inner edge of a thin accretion disc with sub-Eddington luminosities is the inner boundary of the region that most of the luminosity comes from and this happens to be at the Innermost Stable Circular Orbit, ISCO.

In this model, the shear stress is supposed to be a form of viscosity, which is responsible for transporting angular momentum and energy outward and accreting matter inward. Also, it heats up the gas locally. This model introduces viscosity through a so-called $\alpha$-prescription, without specifying the concept of the viscosity itself.

However, according to the very high Reynolds number, the viscosity in the accretion process can not be the same as molecular viscosity and may have the magnetic nature \cite{1991ApJ...376..214B,1992ApJ...400..610B}.

Almost all the accretion disc models assume the dimensionless parameter $\alpha$ to be a constant in this range $0.01-0.1$ \cite{RevModPhys.70.1}. However, global magnetohydrodynamic simulations argued that $\alpha$ is a function of $r$.

The standard $\alpha$ viscosity prescription of the Shakura-Sunyaev model is assumed
\begin{align}\label{VS}
S_{\hat{r}\hat{\phi}}=\alpha P,
\end{align} 
where $S_{\hat{r}\hat{\phi}}$ is the only non-vanishing component of viscous (internal) stress energy tensor in the fluid frame and $P$ is pressure. Also, there is another version of this prescription that is commonly used in terms of the vertically averaged sound speed and the vertical half-thickness of the disc, as
\begin{align}
\nu\simeq\alpha c_s H,
\end{align}
which one should take it by cautious.

In an effort to modify the $\alpha$-prescription, in the literature in general, there are two ways, by considering $\alpha$ as a function of radius, or keeping $\alpha$ a constant and multiply this by a factor e.g. \cite{1996MNRAS.281L..21A,f4a0f084261d4399b3ed571587d7f17c,10.1093/mnras/sts185}.

At the first model, the authors assumed within the ISCO, the viscous torque vanishes and material digs into the central object with the constant energy and angular momentum flux in its frame. This is called the zero-torque boundary condition. However, there has been debate on the validity of this assumption. Some theoretical works suggested that a non-zero-torque at the inner boundary can emerge due to a magnetic field e.g. \cite{1999ApJ...515L..73K,1999ApJ...522L..57G}. However, {Paczy{\'n}ski} in 2000, based on the angular momentum conservation equation, and following by the work of Abramowicz \& Kato in 1989, argued that as long as the shear stress is smaller than the pressure, the thin disc always satisfies the zero-torque condition. Later Afshordi \& {Paczy{\'n}ski}, suggested that the torque at the ISCO is an increasing function of the disc thickness and at the inner edge is small. Also, the inner edge of the disc is almost identical to the place of ISCO. Also, this result satisfied by GRMHD simulations e.g. \cite{2008ApJ...676..549S,2010MNRAS.408..752P} and was shown that for a very thin disc, inside the ISCO the viscous dissipation is negligible and the dissipation profile is identical to that predicted by the standard disc model.
In \cite{Balbus_2017} was argued that in the evolution of thin discs, the
vanishing stress boundary condition will be recovered.

Interestingly, this fact motivated the idea that one can estimate the spin of the black hole by measuring this radius. By the continuum fitting technique, this radius is determined from the temperature maximum of the soft X-ray flux \cite{2011MNRAS.414.1183K}.



%

\section{Equations of thin disc models}\label{sec:equation}

In this section, we introduce the structure of thin accretion disc and state the assumptions and equations in terms of $(t, r, \theta, \phi)$ coordinate, in order to be able to focus on the key physics of the problem and to avoid being distracted by technical details. One could use the transformations \eqref{transf1} to rewrite them in prolate spheroidal coordinates easily.

We assume a steady axisymmetric fluid configuration. In these models, all physical quantities depend on the vertical distance from the equatorial plane and the radial distance from the central object. As a result of the geometrically thin assumption, the two-dimensional disc structure can be decoupled to two one dimensional configurations, namely a radial quasi-Keplerian flow and a vertical hydrostatic structure. In this model usually a vertically integrated approach is used, namely, we integrate along with the height of the disc and neglect the z-dependences of the relevant quantities. However, when the accretion rate is large, one should consider the z-dependence of fluid quantities \cite{2007MNRAS.381.1267K}.

There are three fundamental equations that govern the radial structure of the thin disc model. First, the particle number conservation
\begin{equation}\label{restmasscon}
(\rho u^{\mu})_{;\mu}=0 \,,
\end{equation}
where $u^{\mu}$ is the four-velocity of the fluid and $\rho$ is the rest mass density. The mass accretion rate is connected to this conservation law, meaning we expect the mass accretion rate to be constant, otherwise we would see matter pile up at some certain region of the disc. The other equations are described by the radial component of conservation of energy-momentum tensor $T^{\mu \nu}{}_{;\nu}=0$, parallel to the four-velocity,

\begin{equation}\label{energycon}
u_{\mu} T^{\mu \nu}{}_{;\nu}=0\,.
\end{equation}
And the radial component of projection of this conservation on to the surface normal to the four velocity,

\begin{equation}\label{NSE}
h_{\mu \sigma}(T^{\sigma \nu})_{;\nu}=0 \,,
\end{equation}
where $h^{\mu \nu} = u^{\mu} u^{\nu} + g^{\mu \nu}$ is the projection tensor giving the induced metric normal to $u_{\mu}$. The stress-energy tensor $T^{\sigma \nu}$ reads as,

\begin{align}
T^{\mu\nu}=hu^{\mu}u^{\nu}-pg^{\mu\nu}+q^{\mu} u^{\nu}+q^{\nu} u^{\mu}+S^{\mu\nu},
\end{align}
where $h$ is enthalpy density, which is the sum of internal energy per unit proper volume and pressure over rest-mass density, $p$ is the pressure, $u^{\mu}$ is four-velocity of fluid, $q^{\nu}$ is transverse energy flux and $S^{\mu\nu}$ is the viscous stress-energy tensor. In relativistic form, when we have no bulk viscosity, it is given by $S^{\mu\nu}=-2\lambda \sigma^{\mu\nu}$, where $\lambda$ is the dynamical viscosity and $\sigma^{\mu\nu}$ is the shear tensor, where the only non-vanishing component according to assumptions of the thin disc model is

\begin{align}\label{shearsigmahat}
\sigma_{r\phi}=\frac{1}{2}(u_{r;\beta}h^{\beta}{}_{\phi}+u_{\phi;\beta}h^{\beta}{}_{r})-\frac{1}{3}h{}_{r\phi}u^{\beta}{}_{;\beta}.
\end{align}
The shear rate is a measure of rate of change of the angular velocity with the radius. However, the shear rate is the local measurements and in the model we should consider, this component measured in the fluid frame $\sigma_{\hat{r}\hat{\phi}}$. Indeed, this frame is a proper frame to do minimal modifications to an inviscid flow and contains orthogonal or nonholonomic basis with respect to the chosen metric. This basis is with respect to the local Lorentz frame and used by a real observer versus the holonomic basis represents the global space-time and apart from Cartesian one, they are not orthogonal. In general, one can obtain the nonholonomic basis from the holonomic one by applying the Gram-Schmidt process. The relation between $\sigma_{\hat{r}\hat{\phi}}$ and $\sigma_{r\phi}$ is given by

\begin{align}
\sigma_{\hat{r}\hat{\phi}}=e^{r}{}_{\hat{r}}e^{\phi}{}_{\hat{\phi}}\sigma_{r\phi},
\end{align}
where the basis $e^{\nu}{}_{\hat{\nu}}$ contains orthonormal vectors. This basis is scaled by the coefficient in the original metric to obtain a unit vector. 


By considering the assumption of accretion disc that lies in equatorial plane, one can impose more simplification and ignore terms quadratic in $u^r$ in the shear rate and finish up with a formula for the Kerr metric \cite[equation (5.4.6)]{1973blho.conf..343N}, which is similar to the classical one up to a coefficient.

By applying the assumptions of the thin disc model on the basic equations \eqref{restmasscon}-\eqref{NSE} together with the relations describing radiative energy transport and vertical pressure gradient, we end up with a system of nonlinear algebraic equations governing this disc model \cite{1973blho.conf..343N}, as follows.

The surface density $\Sigma$, is obtained by vertical integration of the density,

\begin{align}\label{sigma2}
\Sigma=\int^{+H}_{-H}\rho {\rm d}z =  2\rho H,
\end{align}
where $H$ is disc height or half of the thickness of the disc.

For steady accretion through the thin discs, it is often assumed that the mass accretion rate is determined by the boundary conditions at a large distance from the central object. Considering this assumption and after vertical and radial integration of the continuity equation, mass accretion rate is given by,

\begin{align}
\dot{M}=-2\pi r u^r \Sigma=\text{constant},    
\end{align}
where $u^r$ is the radial velocity of inflow. However, under some circumstances, the expression for $\dot{M}$ may be not that simple \cite{2001ApJ...547L.151L}. So, the radial velocity of the fluid which is responsible for accreting mass is obtained in terms of the mass accretion rate $\dot{M}$,

\begin{align}\label{massrate}
u^r=-\frac{\dot{M}}{2\pi r \Sigma}.
\end{align}
Although in the disc the motion is nearly circular, by passing the ISCO the radial velocity increases rapidly.


Following the assumption of thin disc models, the heat flow was assumed to be in the vertical direction, meaning $q^z$. Therefore, the time-averaged flux of radiant energy (energy per unit proper area and proper time) flowing out of upper and lower surfaces $F$ relate to the heat flow as \cite{1974ApJ...191..499P,1973blho.conf..343N}, 

\begin{align}
q^z(r,z) = F(r) \frac{z}{H(r)}.
\end{align}
By using the fundamental equations \eqref{restmasscon},\eqref{NSE},\eqref{energycon} and usual manipulation with assumptions we obtain,
\begin{align}\label{ene}
\frac{(\Omega {L}-E)^2}{\Omega_{,{r}}}\frac{{F{r}}}{{\dot{M}}}= \int_{{r}_0}^{r} \frac{(\Omega {L}-{E})}{4\pi}{L}_{,{r}} d{r}, 
\end{align}
where $E$ and $L$ are the energy and angular momentum per unit mass of geodesic circular motion in the equatorial plane, and $\Omega$ is the corresponding angular velocity \cite{2017MNRAS.468.4351C}.

The vertically integrated viscous stress $W$, whose obtained by vertically integration of the viscous stress $S^{\hat{r}\hat{\phi}}$ equation \eqref{VS} is given by 

\begin{equation}\label{w}
W=\int^{+H}_{-H}S{}^{\hat{r}\hat{\phi}} {\rm d}z = 2{\alpha}P {H}.
\end{equation}
The vertically integration generated energy flux via viscosity reads as

\begin{align}\label{navi}
F = -\int^{+H}_{-H} \sigma_{\hat{r}\hat{\phi}}S{}^{\hat{r}\hat{\phi}} {\rm d}z = -{\sigma_{\hat{r}\hat{\phi}}} {W},
\end{align}
where according to the models' assumptions, at each radius the emission is like a black body radiation. The energy transportation law is given by,

\begin{equation}\label{OD}
a T^4={\Sigma} {F}{\kappa},
\end{equation}
where $\kappa$ is Rosseland-mean opacity,
\begin{align}
\kappa=0.40 + 0.64\times10^{23} (\frac{\rho}{{\rm g} \quad cm^{-3}})(\frac{T}{K})^{-\frac{7}{2}} cm^2{\rm g}^{-1},
\end{align}
where the first term is electron scattering opacity and the second one is free-free absorption opacity. And $a$ is the radiation density constant giving by

\begin{align}
a= \frac{4 \sigma}{c},
\end{align}
where $\sigma$ is Stefan-Boltzmann’s constant.


In general, for the vertical direction in the comoving frame of fluid, the force due to vertical pressure gradient is in the balanced with gravity, the centrifugal force and vertically Euler force \footnote{Of course, in this frame the Coriolis force vanishes.}. The pressure $P$, is the sum of gas pressure from nuclei and the radiation pressure,

\begin{equation}\label{P}
P=\frac{{\rho kT}}{{m_p}}+\frac{{a}}{{3}}{T}^4,
\end{equation}
where ${m_p}$ is the rest mass of the proton, ${k}$ is Boltzmann's constant, $a$ is the radiation density constant, and $T$ is the temperature. In fact, we ignore the mass difference between neutrons and protons in the first term for simplicity. In practice, the pressure equation in the vertical direction is given by

\begin{equation}\label{VP}
\frac{P}{{\rho}}=\frac{1}{2}\frac{{(HL)^2}}{{r^4}}\,,
\end{equation}
which is derived from the relativistic Euler equation with no additional simplifications \cite[equation 28]{1997ApJ...479..179A}.


By solving these eight equations \eqref{sigma2}, \eqref{massrate}, \eqref{ene}, \eqref{w}, \eqref{navi}, \eqref{OD}, \eqref{P} and \eqref{VP}, one obtains the radial profiles of the eight variables in the model, namely the half-thickness of the disc over radius $\frac{H}{r}=h$, surface density $\Sigma$, the central temperature $T$, pressure $P$, the radial velocity $u^r$, the radiation flux $F$, viscous stress $W$ and $\rho$ and from them one can calculate other physical quantities subsequently. All these parameters are functions of the distance from the central object. Also, there are five parameters describing a thin disc solution in this space-time, namely $M$, mass accretion rate $\dot{M}$, $\alpha$, quadrupoles $\rm p$ and $\rm q$, whose we specified in section \ref{sec:results} for plots. 


\section{Construction the thin accretion disc around the distorted naked singularity}\label{sec:eq}

\begin{figure}
\includegraphics[width=0.64\textwidth]{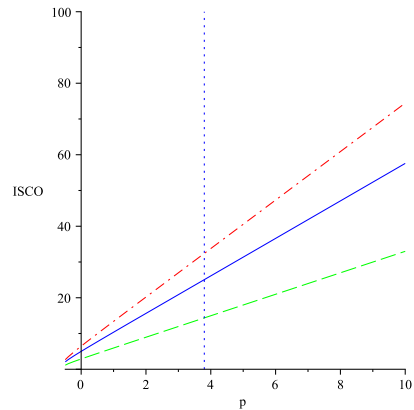}
   \caption{\label{ISCOS}The solid line is places of ISCO for $\rm q=0$, dot-dashed line is places of ISCO for $\rm q_{\rm max}$, and dashed line is for $\rm q_{\rm min}$. In all cases, ISCO is a monotonically increasing function of $\rm p$. Also, any vertical line, $\rm p$=constant, intersect with all lines, where shows the place of ISCO at a fixed $\rm p$, for $\rm q<0$, $\rm q=0$ and $\rm q>0$ simultaneously. So, also the ISCO is a monotonically increasing function of $\rm q$.}
\end{figure}

First of all, we need to rewrite the metric \eqref{EImetric}, in the equatorial plane, meaning $\theta =\frac{\pi}{2}$ or equivalently ${y}=0$ in the coordinates \eqref{transf1},

\begin{align}\label{EIeq}
	ds^2 &= - \left( \frac{x-1}{x+1} \right)^{(1+{\rm p})} e^{2\hat{\psi}} d t^2+ M^2(x^2-1) e^{-2\hat{\psi}} \nonumber\\
	 &\left( \frac{x+1}{x-1} \right)^{(1+{\rm p})}\left[ \left(\frac{x^2-1}{x^2}\right)^{{\rm p}(2+{\rm p})}e^{2\hat{\gamma}}\right.\\
	 &\left. \left( \frac{d x^2}{x^2-1}+d y^2 \right)+ \dd{\phi}^2\right].\
\end{align}
 In the equatorial plane $y=0$, the distortion functions,  up to the quadrupole moment  \eqref{psigeneralequ}-\eqref{gammageneralequ}, simplify to
 
\begin{align}\label{gammapsi1}
	\hat{\psi} & = -\frac{\rm q}{2}({x}^2-1)\,, \\
\label{gammapsi2}
	\hat{\gamma} & = -2{\rm q}{x}+\frac{{\rm q}^2}{4}({x}^2-1)^2\,.
\end{align}
Second step is to determine the inner edge of the disc. As it was mentioned earlier the inner edge of the standard thin disc model is assumed to be at the innermost stable circular orbit. So, we need to analyze the location of the ISCO in this solution. The circular orbits in equatorial plane in this space-time was studied in \cite{2020arXiv201015723F} by means of analysing the effective potential in the equatorial plane,

\begin{align}
    V_{\rm Eff}=& \left( \frac{x-1}{x+1} \right)^{({\rm p}+1)} e^{2\hat{\psi}}\nonumber\\
    &\left[ 1+\frac{L^2e^{2\hat{\psi}}}{M^2(x+1)^2} \left( \frac{x-1}{x+1} \right)^{{\rm p}}\right].\,
\end{align}
It was shown that there is a valid range for $\rm q$, quadrupole moment of source, for each choice of $\rm p$, quadrupole of the central object, for time-like trajectories. First the domain of the central object quadrupole is $\rm p\in[-1+\frac{\sqrt{5}}{5},\infty)$. Also, for each choice of $\rm p$, there exist ISCOs for time-like geodesics for some range of quadrupole of external matter $\rm q\in[\rm q_{min}, q_{max}]$, such that in general, $\rm q_{min}\approx -2.5\times10^{-1}$ at $\rm p=-0.5$ and $\rm q_{max}\approx 4.1\times10^{-3}$ at the minimum of $\rm p$, meaning $\rm p=-0.5528$. Also, from $\rm p=-0.5$, $|\rm q|$ is a monotonically decreasing function of $\rm p$ and approaches to zero. Some of the examples are listed in table \ref{T1}. Also, the place of ISCO in general is closer to the horizon for negative quadrupoles and it is farther away from it for positive quadrupoles in both cases, see Figure \ref{ISCOS}. Therefore, from an ISCO analysis we got a restriction on choosing the quadrupole values in the model.

Indeed, one can derive the various physical quantities appearing in the equations of the thin disc model described earlier in section \ref{sec:equation}. In fact, by its definition the distorted metric is only valid locally. We can see this in the analyze of the behavior of physical quantities like $E=-u_t$, $L=u_\phi$ and $\Omega=\frac{u^{\phi}}{u^t}$ in this space-time, and consider the valid region where they are real-valued, for a given values of quadrupoles. Besides, the angular velocity, which is supposed to be a monotonically decreasing function for vanishing quadrupoles, in this space-time may possess an extremum at some $x$. A likely explanation is that from this point the effect due to the external matter starts to reveal. So, one can take this as the signature of external matter indicating that from this distance $x$, the local solution is no longer valid.

\begin{figure}
                      \includegraphics[width=0.64\textwidth]{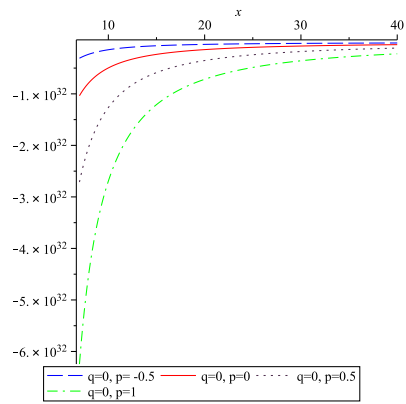}
    \caption{\label{0004} Shear rate for any chosen value of $\rm p$, is negative when there is no external source.}
\end{figure}

\begin{figure}
                  \includegraphics[width=0.64\textwidth]{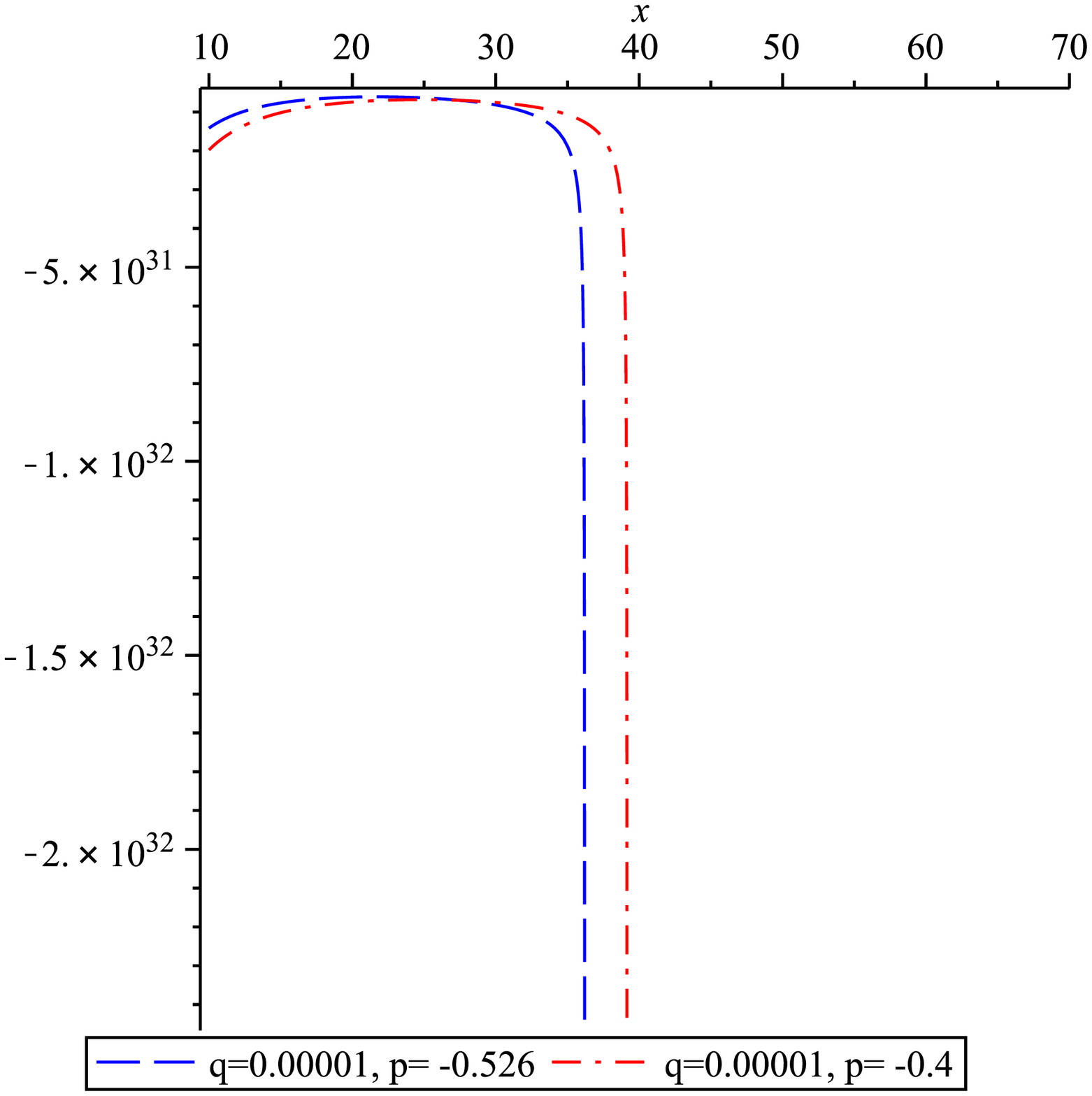}
         \includegraphics[width=0.64\textwidth]{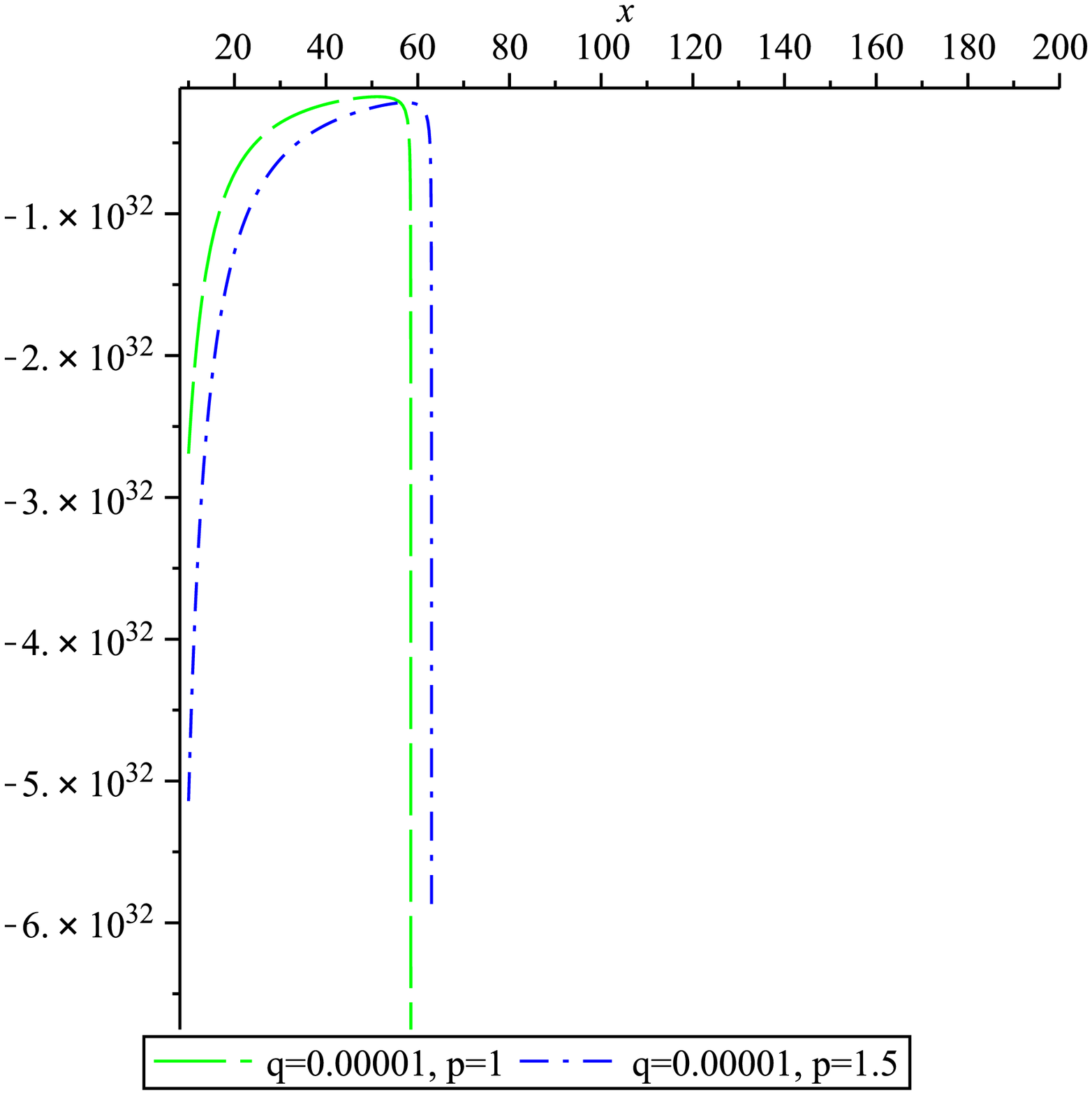}
    \caption{\label{25}For positive values of $\rm q$, always $\sigma_{\hat{x}{\phi}}$ is negative and restricted to a very narrow region.}
\end{figure}

\begin{figure}
\includegraphics[width=0.7\textwidth]{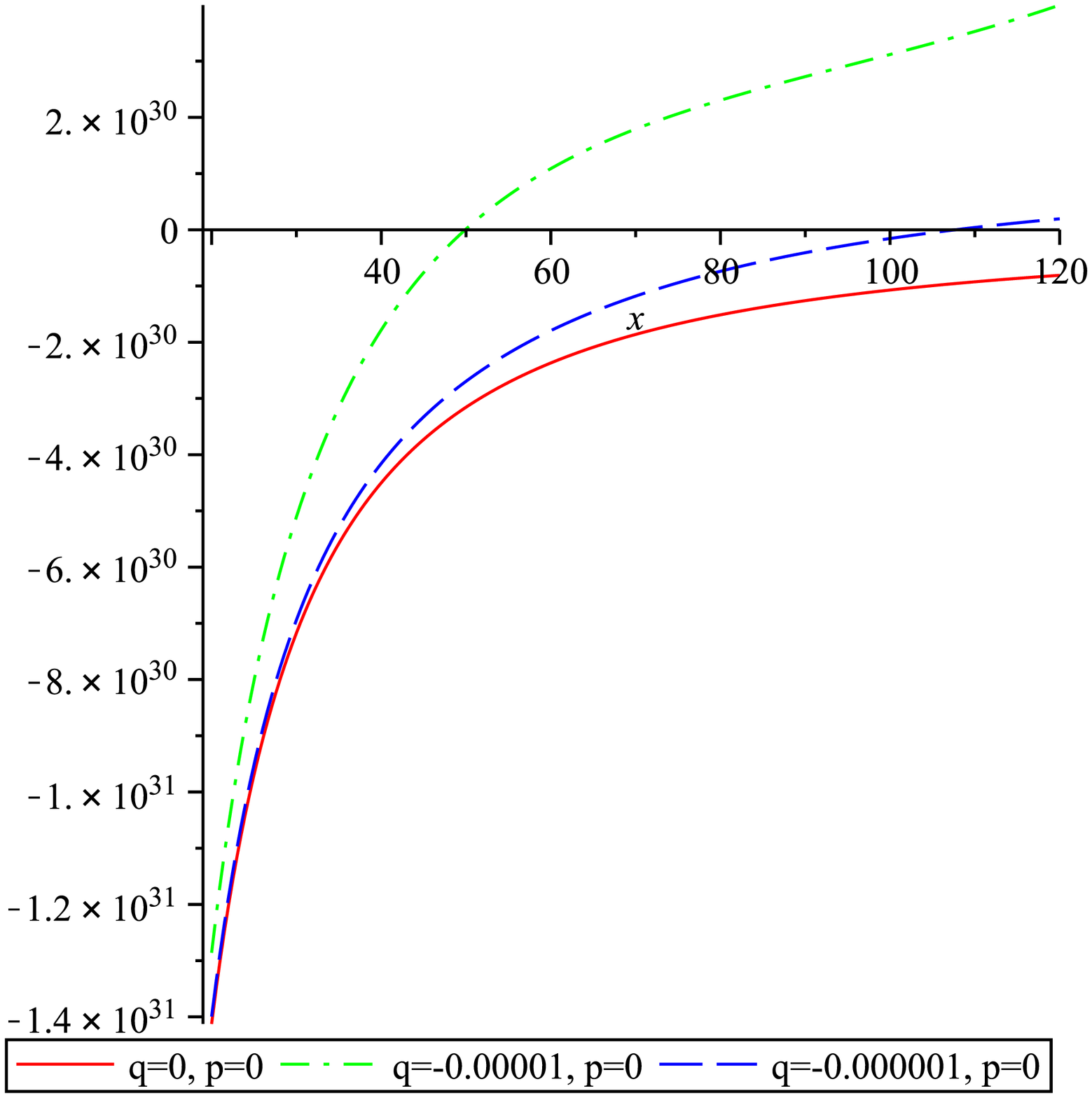}
        \includegraphics[width=0.7\textwidth]{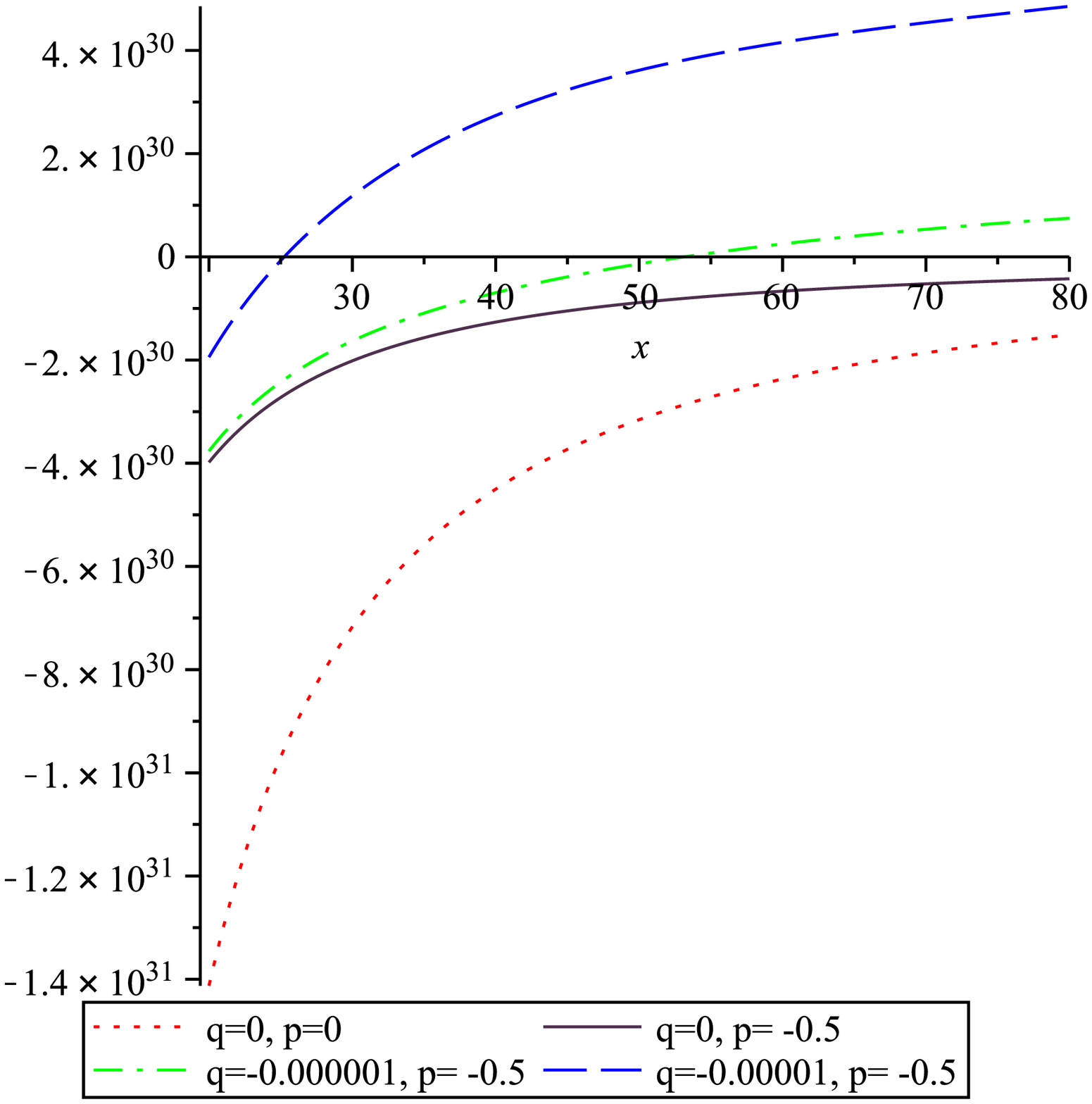}
         \includegraphics[width=0.7\textwidth]{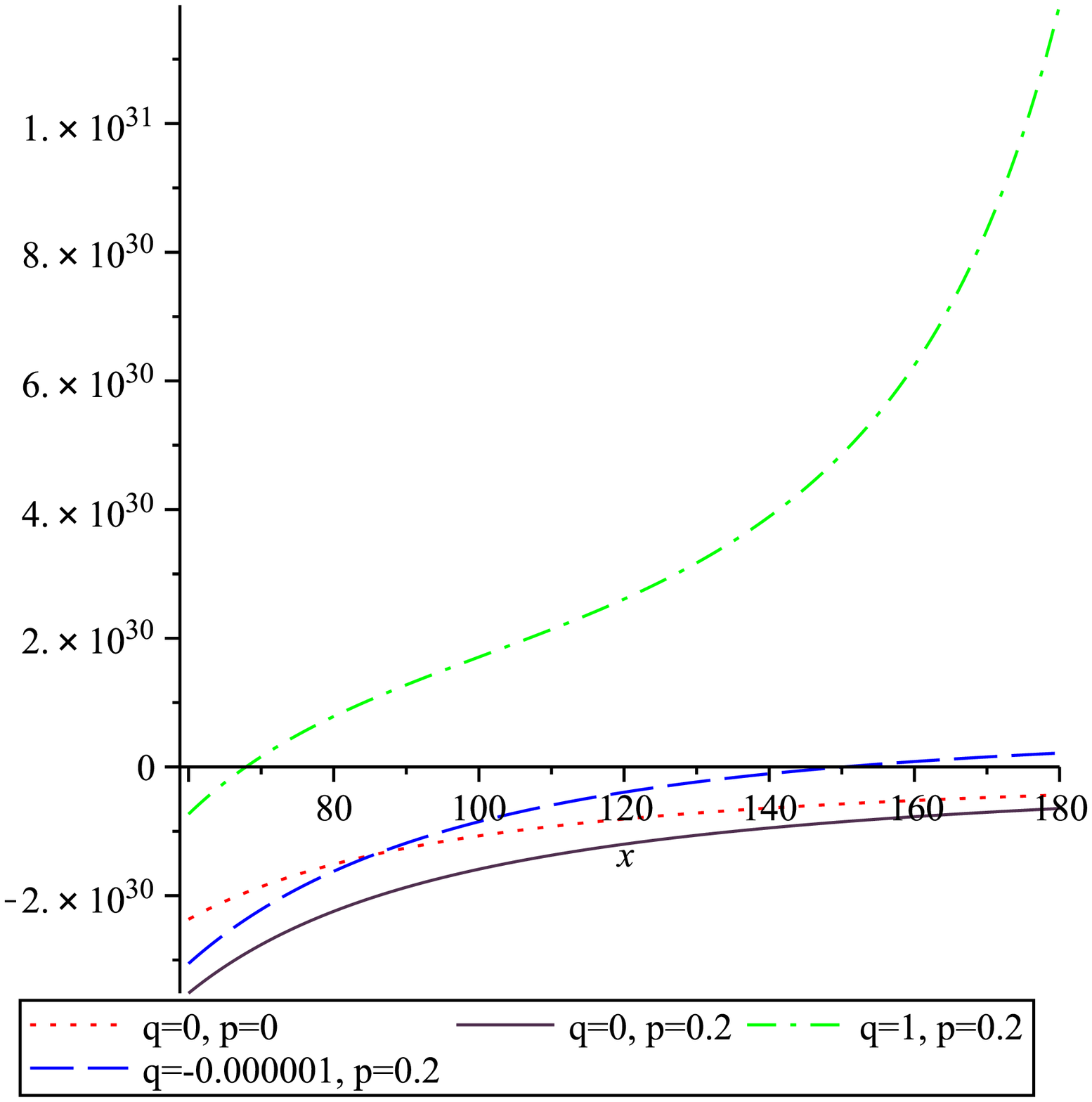}
    \caption{\label{0502} For negative $\rm q$, and any choice of $\rm p$, at some $x_0$, shear rate changes the sign, which means the valid region is restricted to $x_0$.}
\end{figure}

\begin{figure}
               \includegraphics[width=0.7\textwidth]{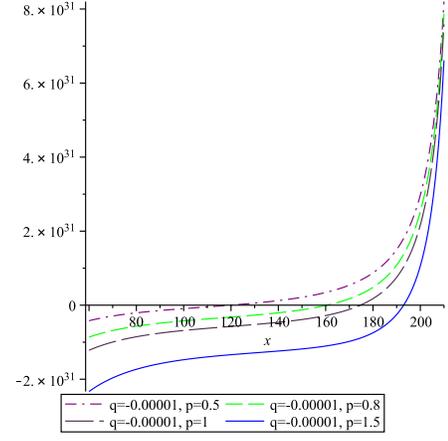}
                   \caption{\label{146810} As $\rm p>0$ gets larger, we see the behaviour of the $\sigma_{\hat{x}\hat{\phi}}$ in the valid region, become different and tends to have small local extermum within this region. Also, as $\rm p$ increases the magnitude of shear rate also increases.}
\end{figure}

Further analysis shows that, although angular velocity gives us an upper bound for the valid region, one can find a better estimate on the valid region by analysing the shear rate \eqref{shearsigmahat}, in the local rest frame $\sigma_{\hat{x}\hat{\phi}}$ \footnote{The calculation of $\sigma_{\hat{x}\hat{\phi}}$ contains terms from the projection tensor $h^{\mu \nu}$, the partial derivative of the four velocity, and the Christoffel symbols that for this space-time they are introduced in \cite{2020arXiv201015723F}}. However, this is also a measurement of the rate of change of the angular velocity with the radius. An analysis shows that, for positive values of $\rm q$ these two estimations are in a good agreement. Consequently, for asymptotically flat solutions, meaning $\rm q=0$, always $\sigma_{\hat{x}{\phi}}$ is negative for this space-time, therefore there is no restriction for any chosen value of $\rm p$, see Figure \ref{0004}. However, the place of ISCO, inner edge, will be different as it was discussed earlier. Also, for positive values of $\rm q$, meaning $(\rm q>0, \rm p)$ always $\sigma_{\hat{x}{\phi}}$ is negative; however, it is restricted to some region, see Figure \ref{25}. For negative $\rm q$, meaning $(\rm q<0, \rm p)$ at some point say $x_0$, shear rate $\sigma_{\hat{x}\hat{\phi}}$ changes the sign, which means the valid region is restricted to $x_0$, see Figure \ref{0502}. However, in this case depends on the value of $\rm p$, the behaviour of the $\sigma_{\hat{x}\hat{\phi}}$ in the valid region, is different, for example see Figure \ref{146810}. Since $\sigma_{\hat{x}{\phi}}$ is the major contributing factor mostly in vertical structure, we expect to have some effects, especially on this direction. Nevertheless, this effect becomes more strong for larger values of $\alpha$. We should mention that as the absolute value of the quadrupole $|\rm q|$ approaches to zero, we arrive at a wider valid range for $x$, as it is expected from the $(\rm q=0, \rm p)$ limit for each choice of $\rm p$.

The final result is obtained by intersection the results from analysis of physical quantities being real-valued and from the upper bound given by $\sigma_{\hat{x}\hat{\phi}}$ analysis. This gives us some restrictions on the range of the ${x}$ coordinate for any chosen value of $\rm p$ and $\rm q$.

As the last step, before solving the system of equations in section \ref{sec:equation}, one needs to rewrite equations and quantities in the prolate spheroidal coordinates $(t, x, y, \phi)$ simply via the transformation \eqref{transf1} to obtain the system of algebraic equations that can be treat analytically. Also, this system of equations admits a unique solution for considering positive temperature and pressure. The results are described and plotted in the next section.


\section{Results from the thin disc model}\label{sec:results}

In fact, as the distorted solutions are only valid locally, we only consider the inner part of the disc in this space-time. As it was discussed earlier the upper limit of the solution depends on the choice of both the quadrupoles $\rm p$ and $\rm q$. The results were produced by Mathematica, and the physical constants and parameters that were used are as follows 

\begin{align}\label{param}
G=&6.67 \times 10^{-8} \quad {\rm cm^3 g^{-1} sec^{-2}},\\
c=& 3 \times10^{10} \quad {\rm cm \hspace{0.1cm} sec^{-1}}, \\
\kappa_{es}=& 0.40 \quad {\rm cm^2 g^{-1}},\\
M=& 10^{34} \quad \rm g,\\
\dot{M}=&0.25 \dot{M}_{\rm Edd}(=10 \frac{L_{\rm Edd}}{c^2})\quad {\rm g}\hspace{0.1cm} {\rm sec}^{-1}\label{parame},\\
\alpha=& 0.02.\
\end{align}
Apart from height scale $h$, which is a dimensionless height scale, the results are represented in cgs units.

Let us first discuss the undistorted naked singularities and static black hole, meaning $(\rm q=0, \rm p)$. For any negative value of $\rm p$, plots start at a place closer to the horizon and they get further away as $\rm p$ increases. This is because the inner edge of the disc, ISCO for negative quadrupoles are closer to the horizon and the opposite for positive ones, Figure \ref{ISCOS}. However, the maximum of all plots are much higher for the negative quadrupoles $\rm p$, compared to the Schwarzschild case $\rm p=0$, and to the positive quadrupoles $\rm p$.

\begin{figure}
        \includegraphics[width=0.6\textwidth]{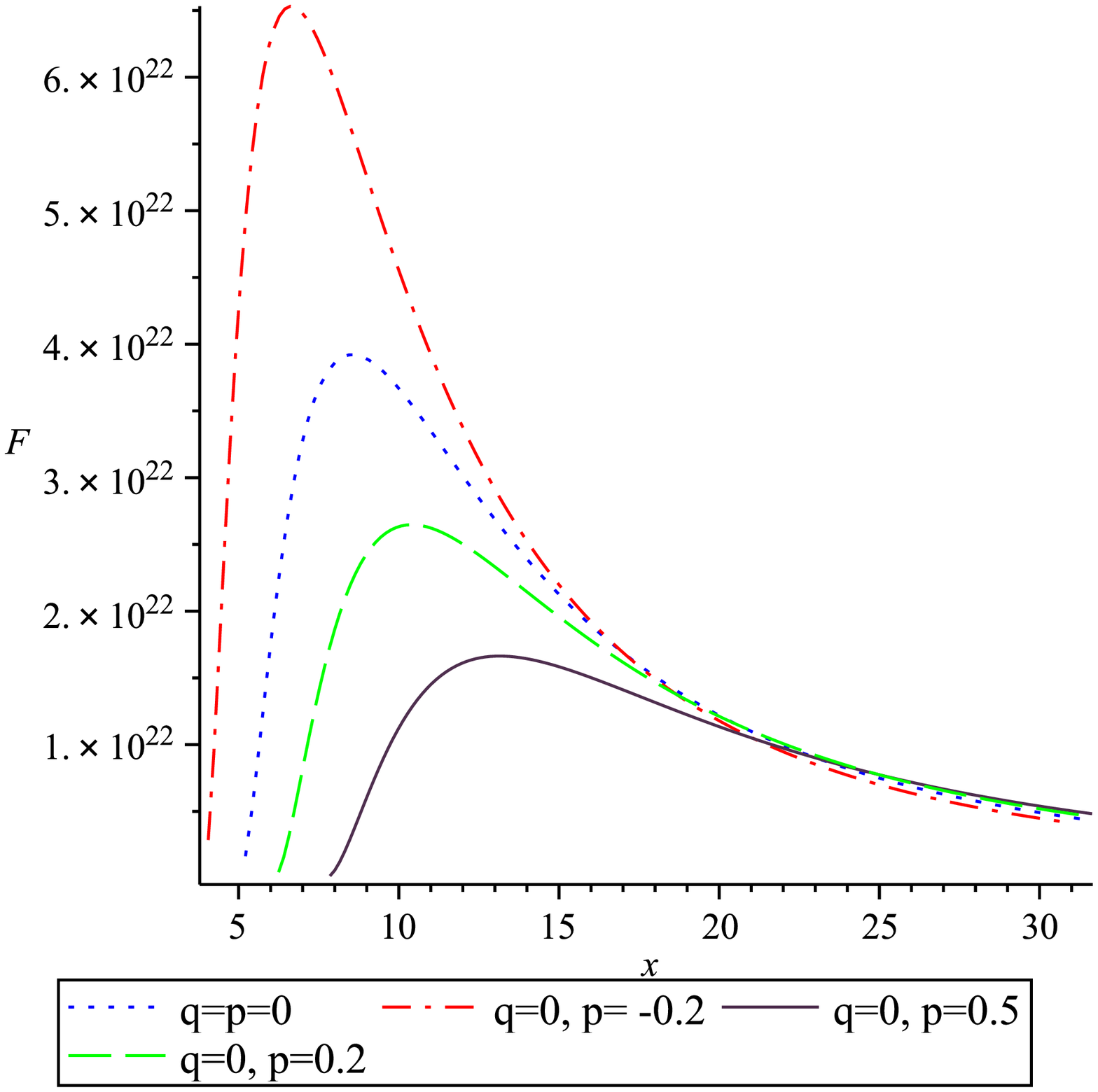}
        \includegraphics[width=0.6\textwidth]{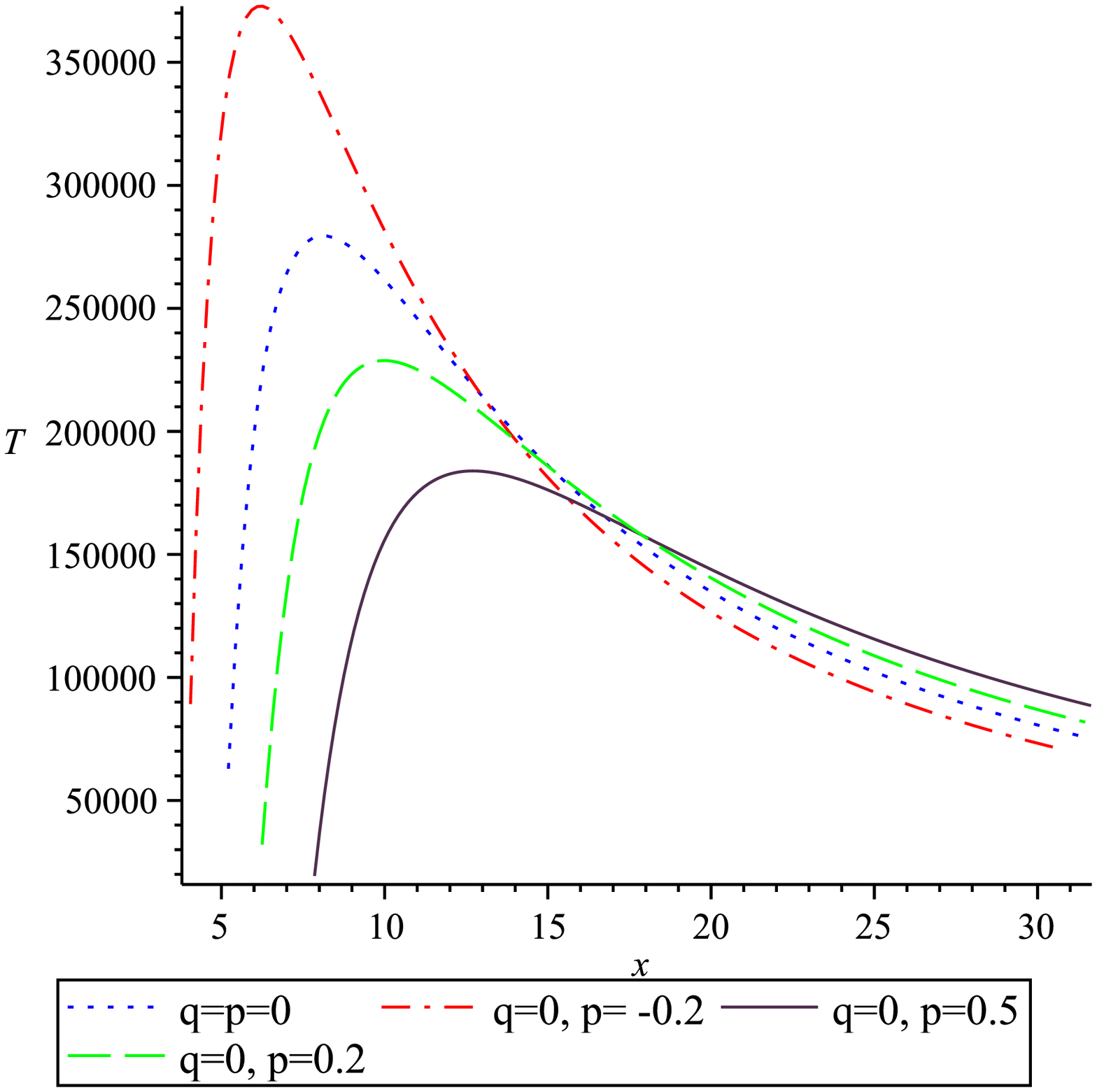}
        \includegraphics[width=0.6\textwidth]{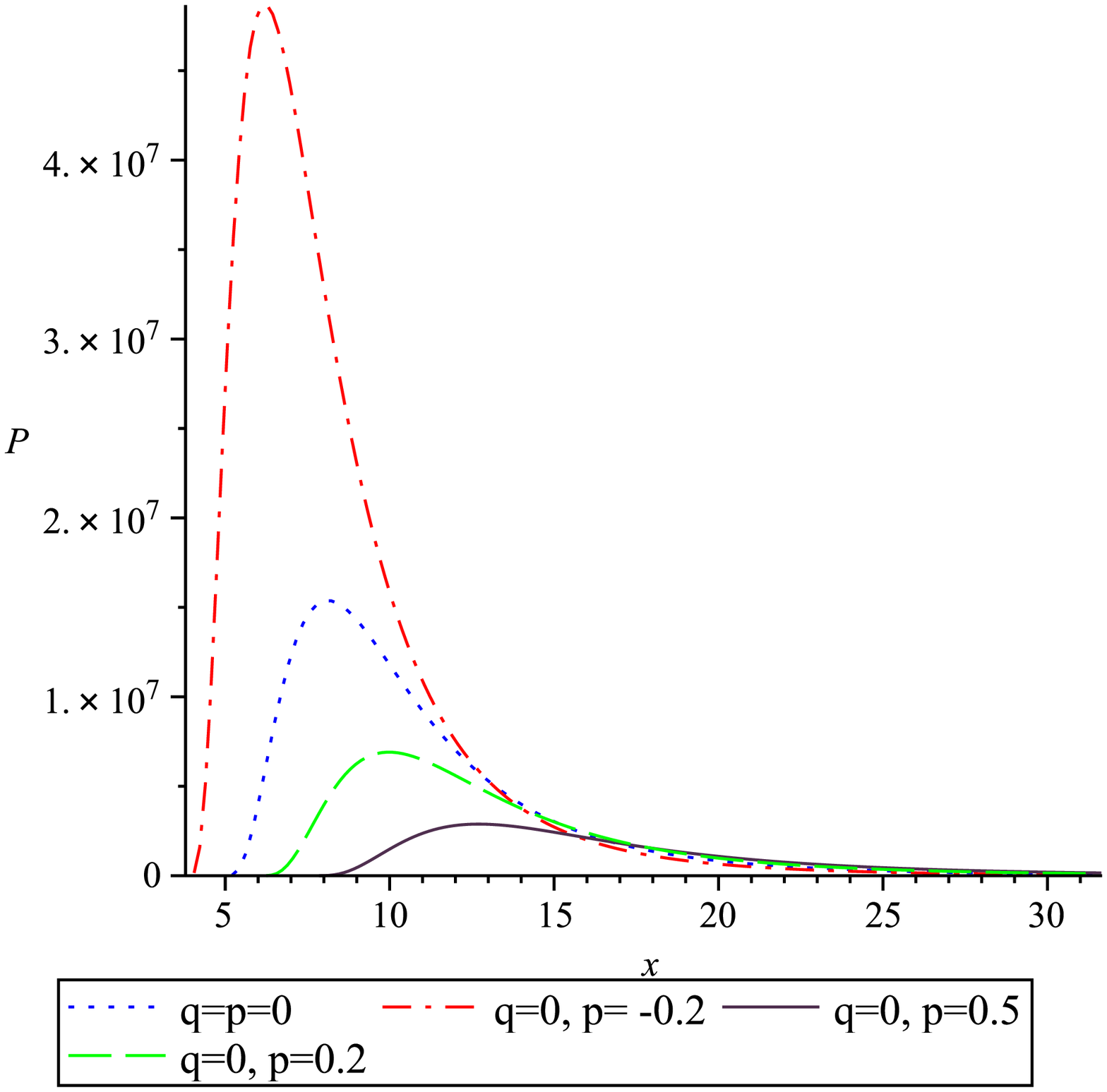}
          \includegraphics[width=0.6\textwidth]{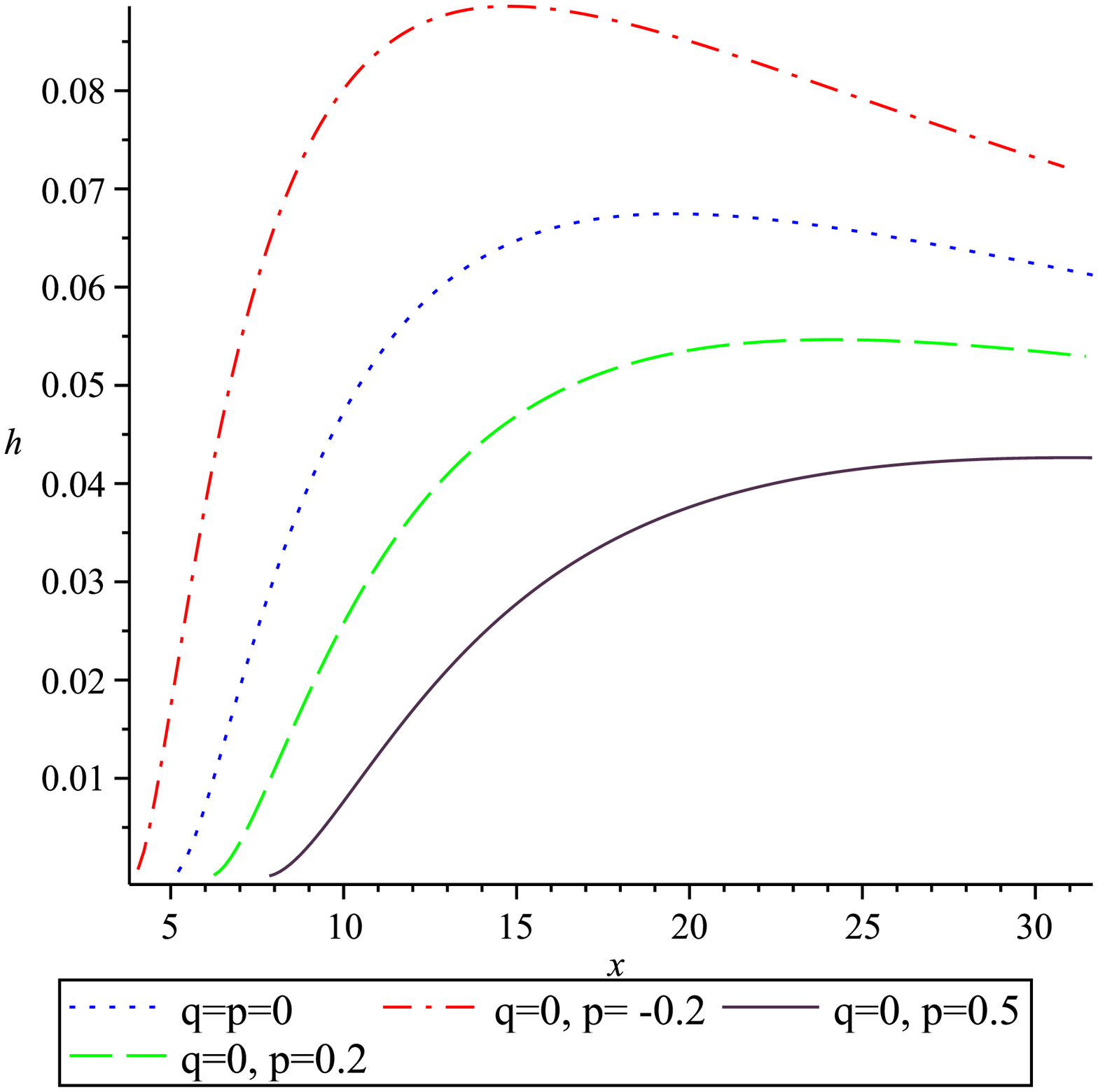}
     \caption{\label{ftponlyp}Radiation flux $F$ $(\rm g/s^3)$, temperature $T$ $(K)$, pressure $P$ $(\rm g/s^2cm)$, and the height scale of the disc $h$.}

       \end{figure}
In fact, as we see in Figure \ref{ftponlyp}, these quantities have almost the same pattern. In particular, in the case of negative quadrupoles $\rm p$, there is a sharper ascent and descent rather than for positive quadrupoles $\rm p$. In this regard, the place of the pick of each plot, is a monotonically decreasing function in $\rm p$. Therefore, there are distinguishable differences between the properties of a standard thin disc around naked singularities and Schwarzschild black hole, in general.

\begin{figure}
    \includegraphics[width=0.525\textwidth]{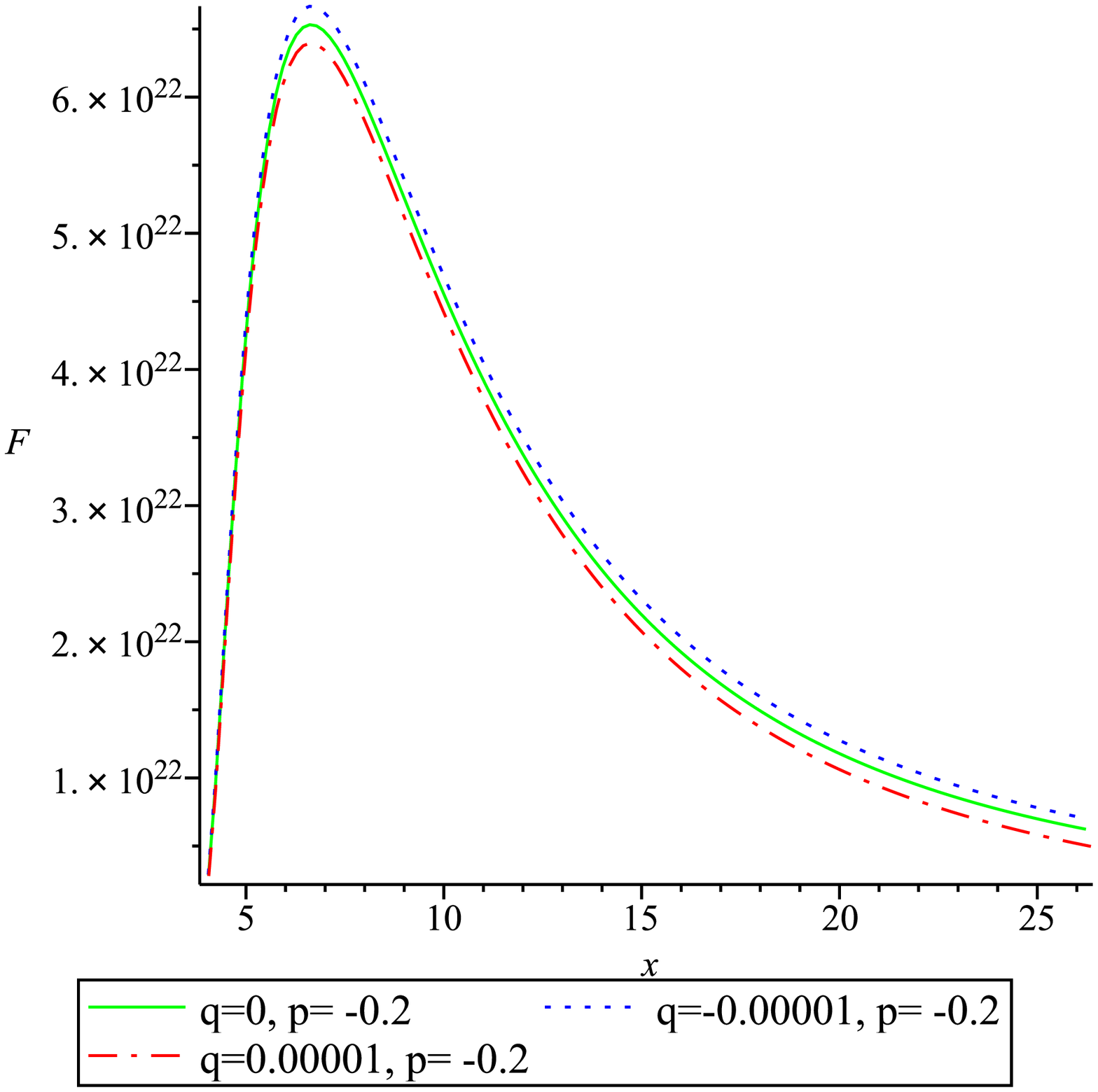}
                \includegraphics[width=0.525\textwidth]{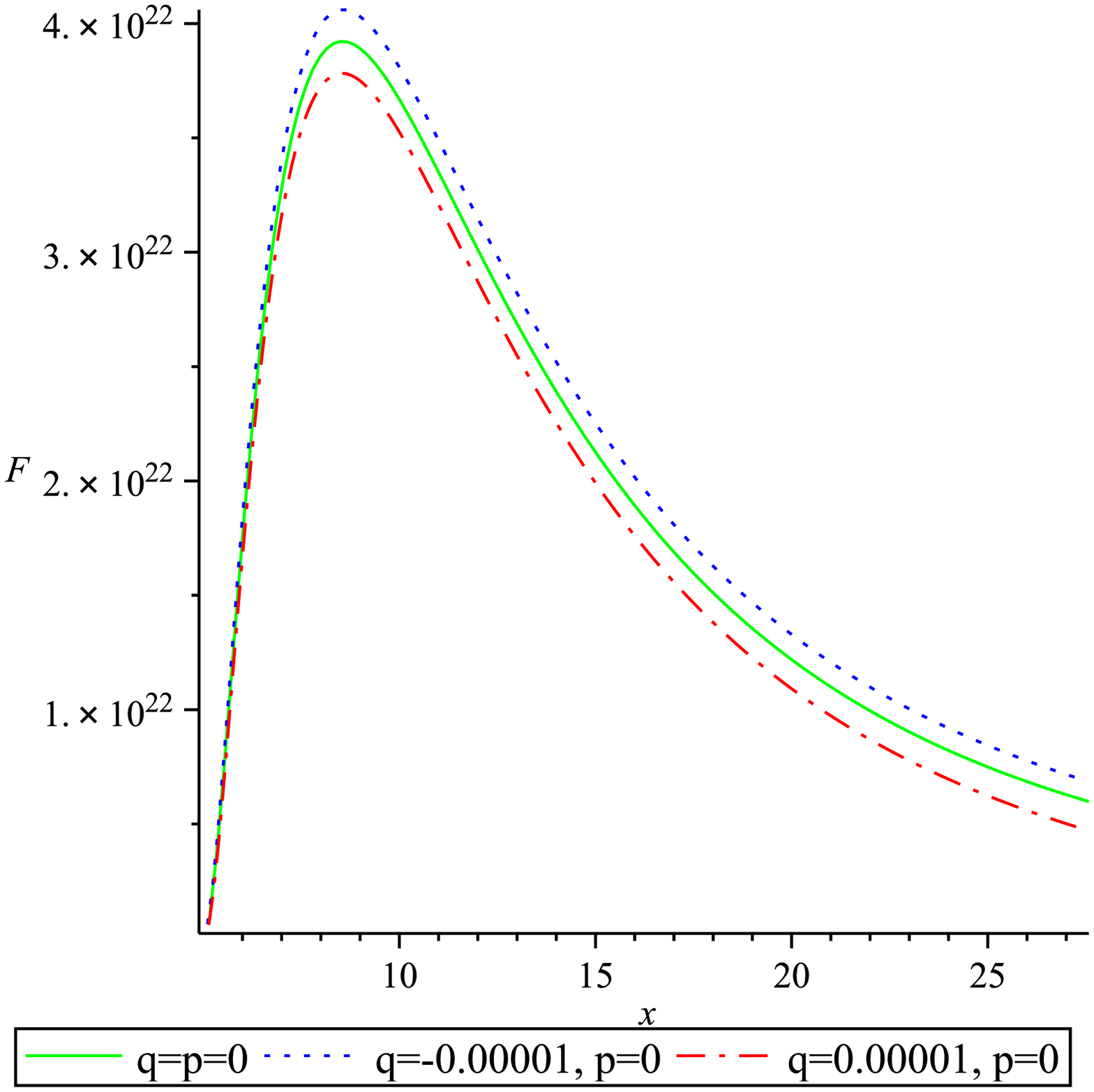}
                 \includegraphics[width=0.525\textwidth]{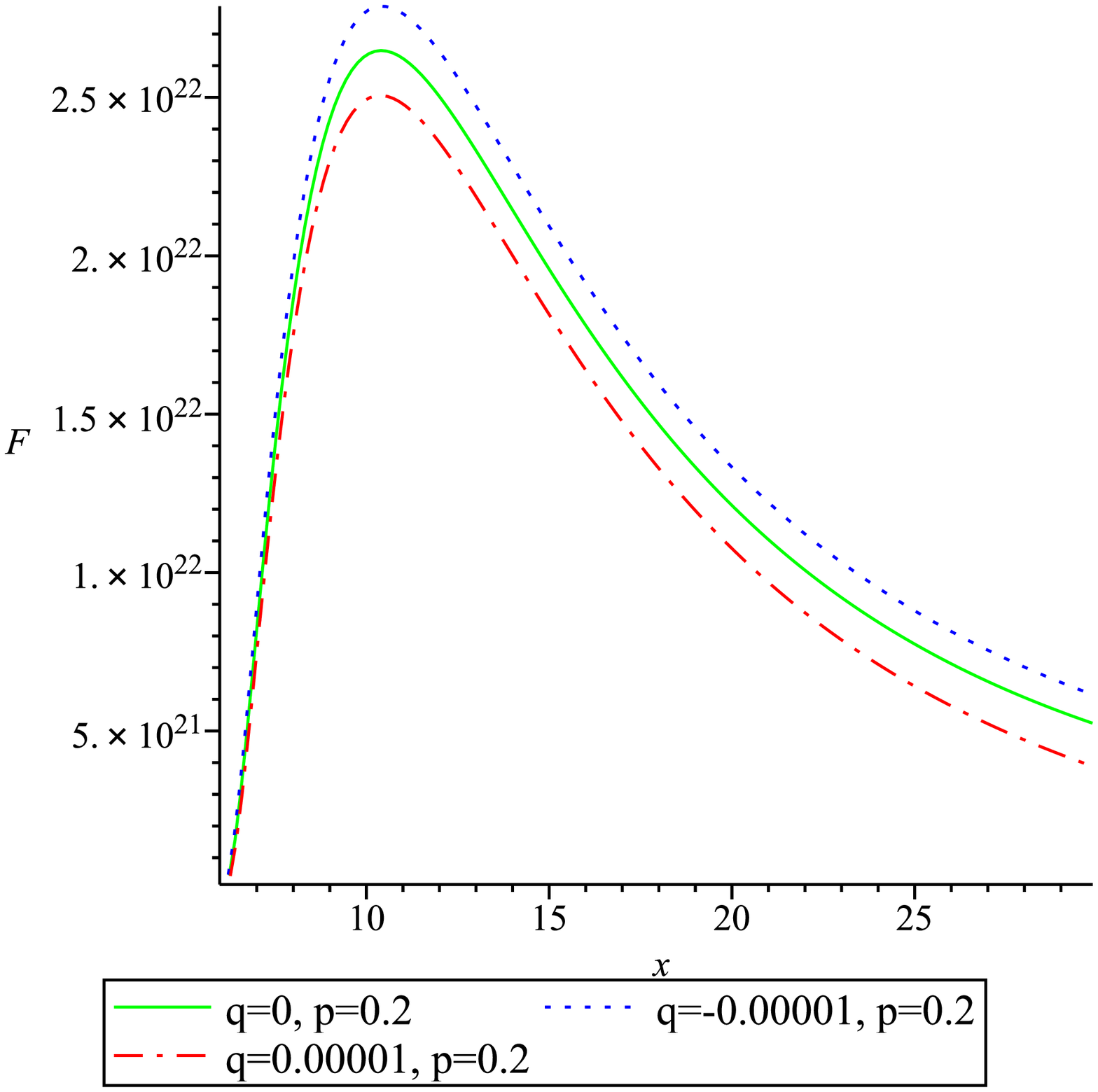}
    \includegraphics[width=0.525\textwidth]{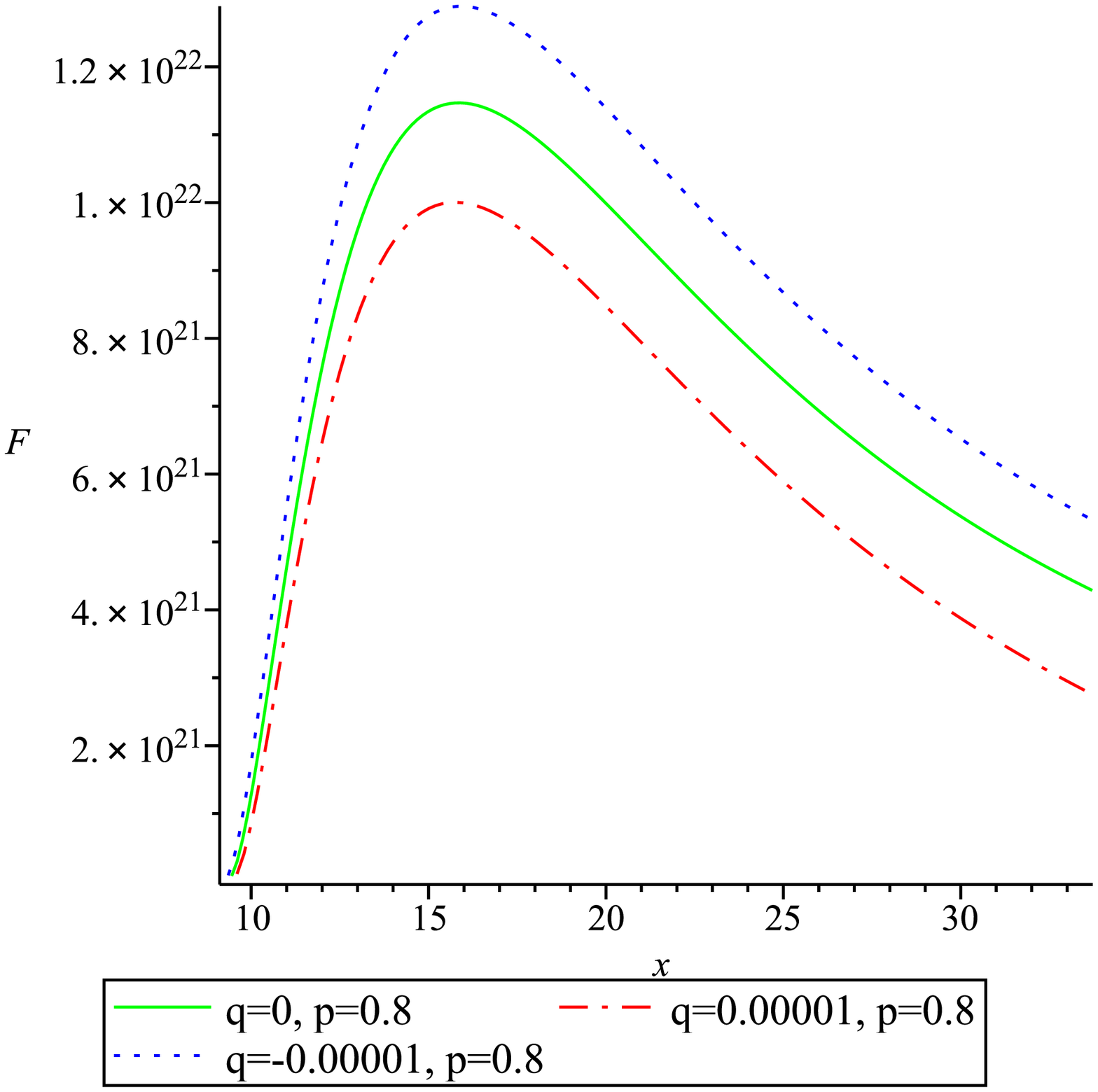}
     \includegraphics[width=0.525\textwidth]{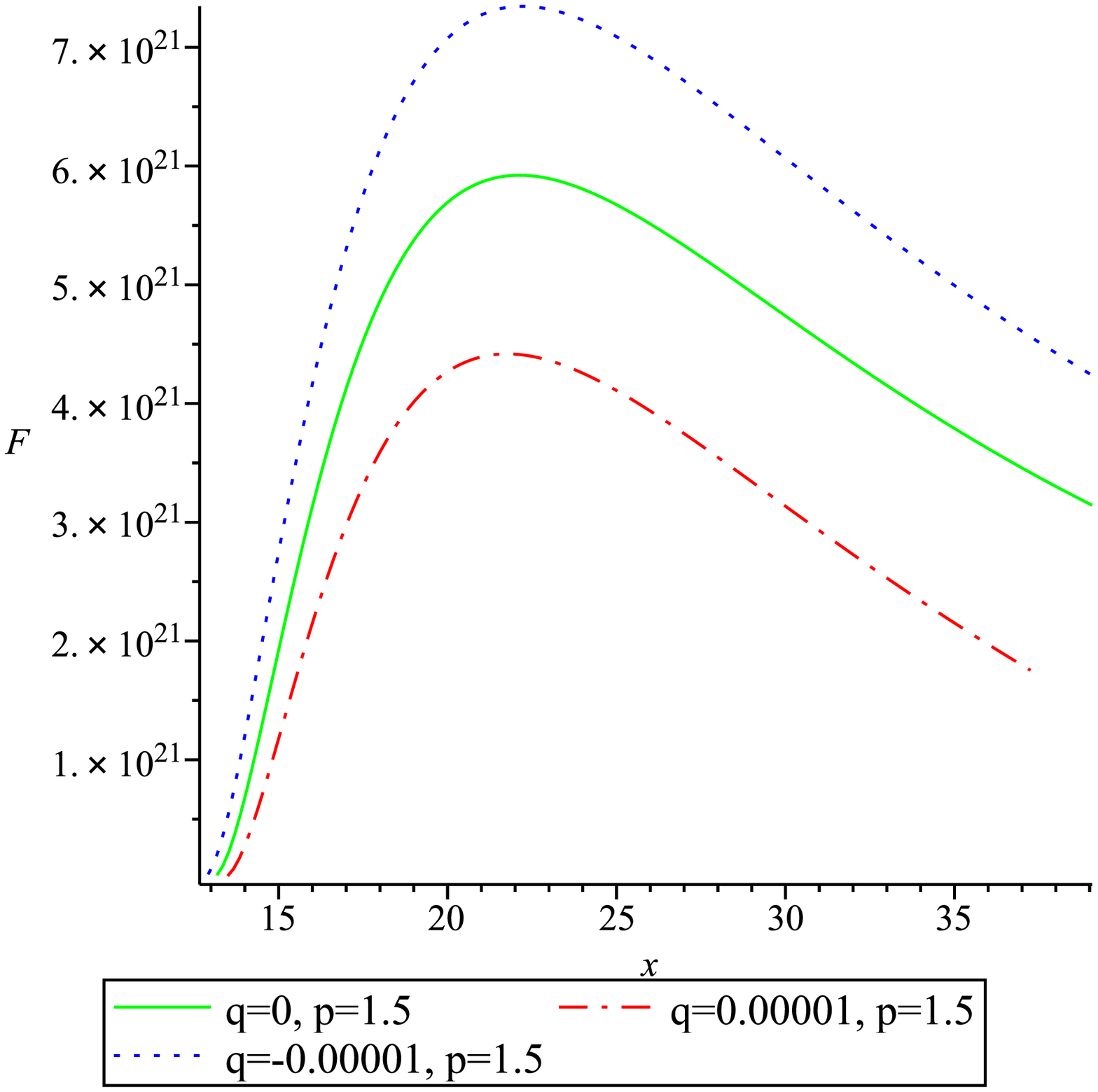}
       \caption{\label{fplot} Radiation flux $F$ $(\rm g/s^3)$ for $\rm p<0$, $\rm p=0$ and $\rm p>0$.}
\end{figure}

\begin{figure}
    \includegraphics[width=0.525\textwidth]{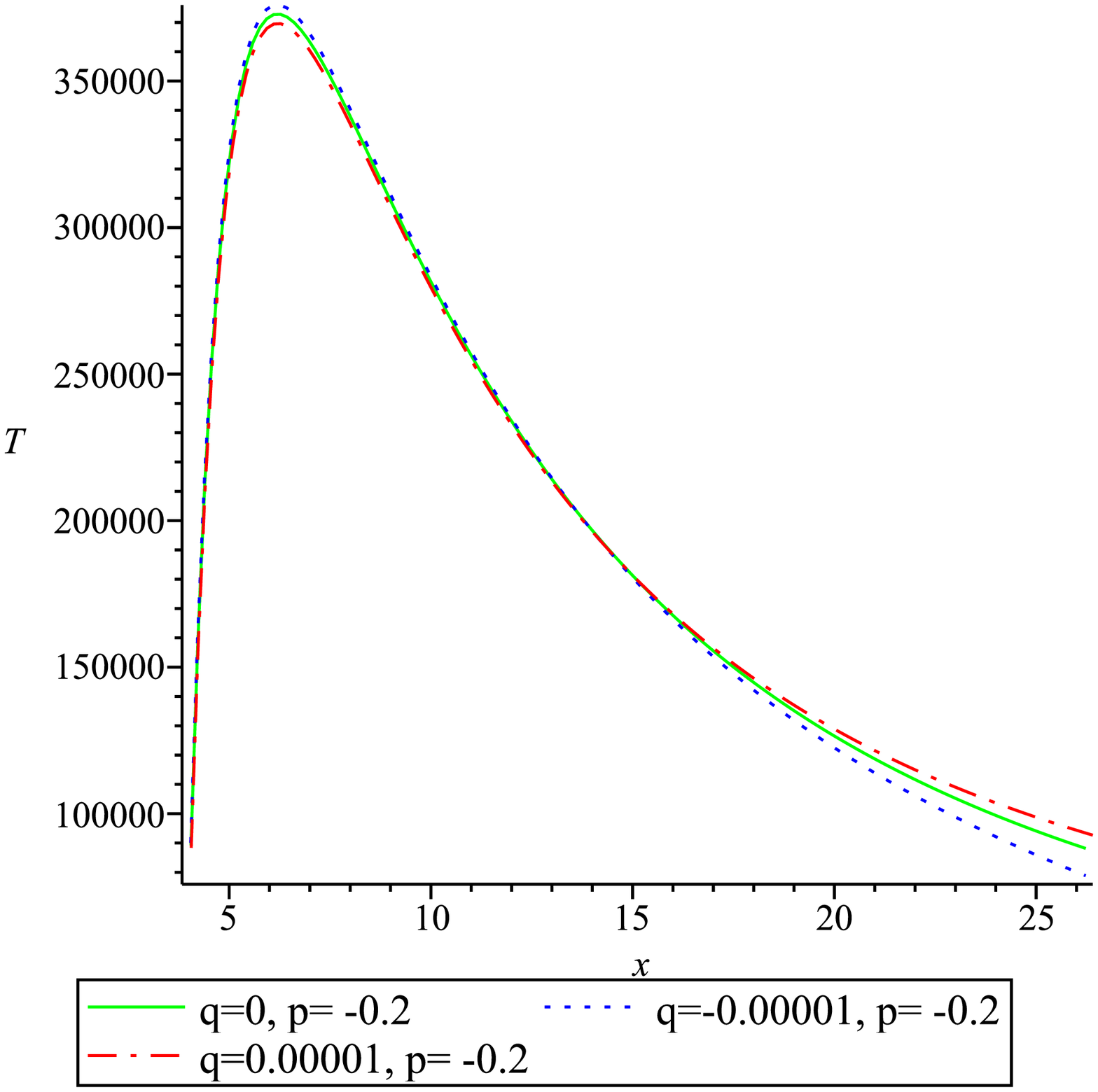}
        \includegraphics[width=0.525\textwidth]{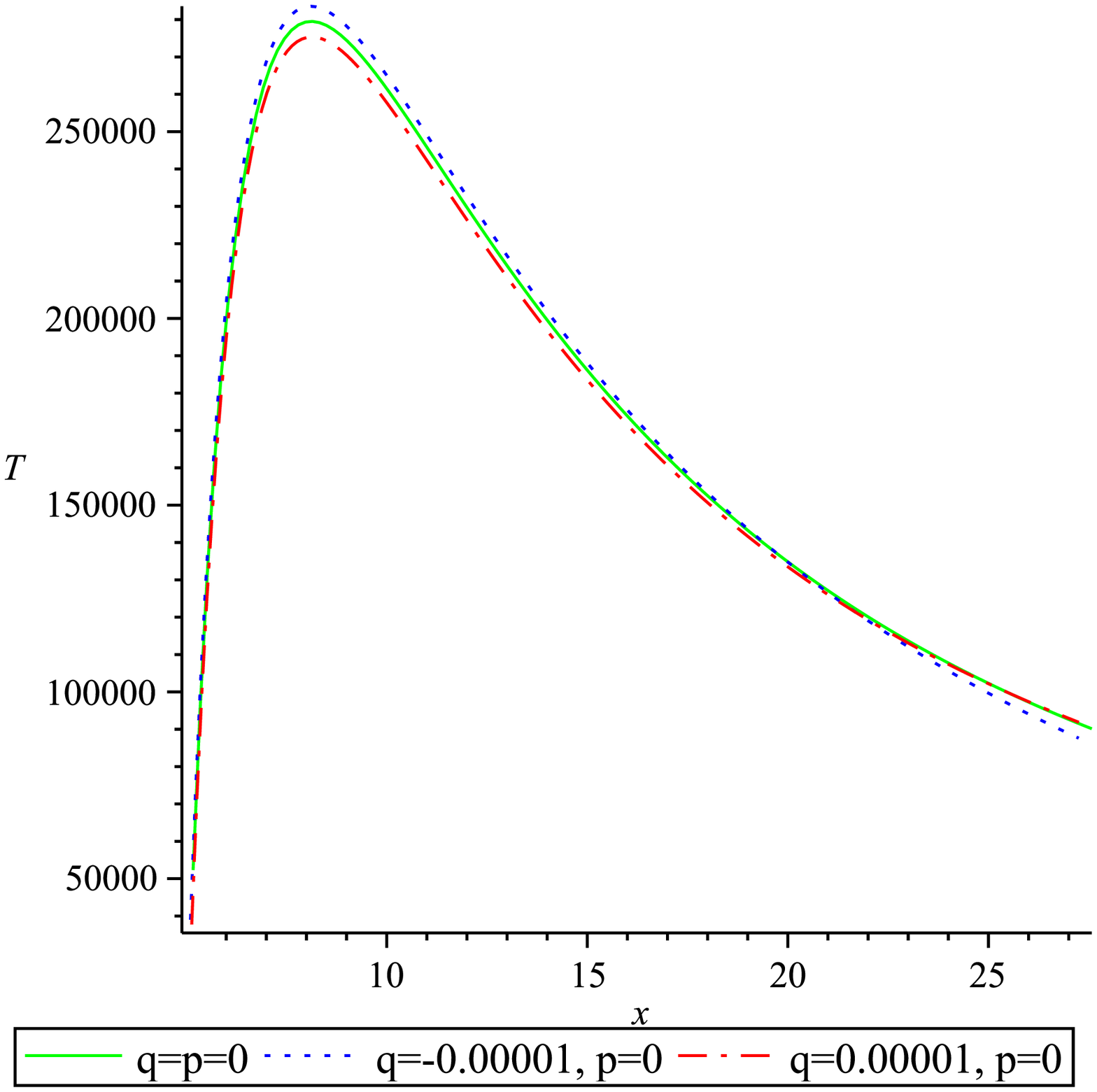}
        \includegraphics[width=0.525\textwidth]{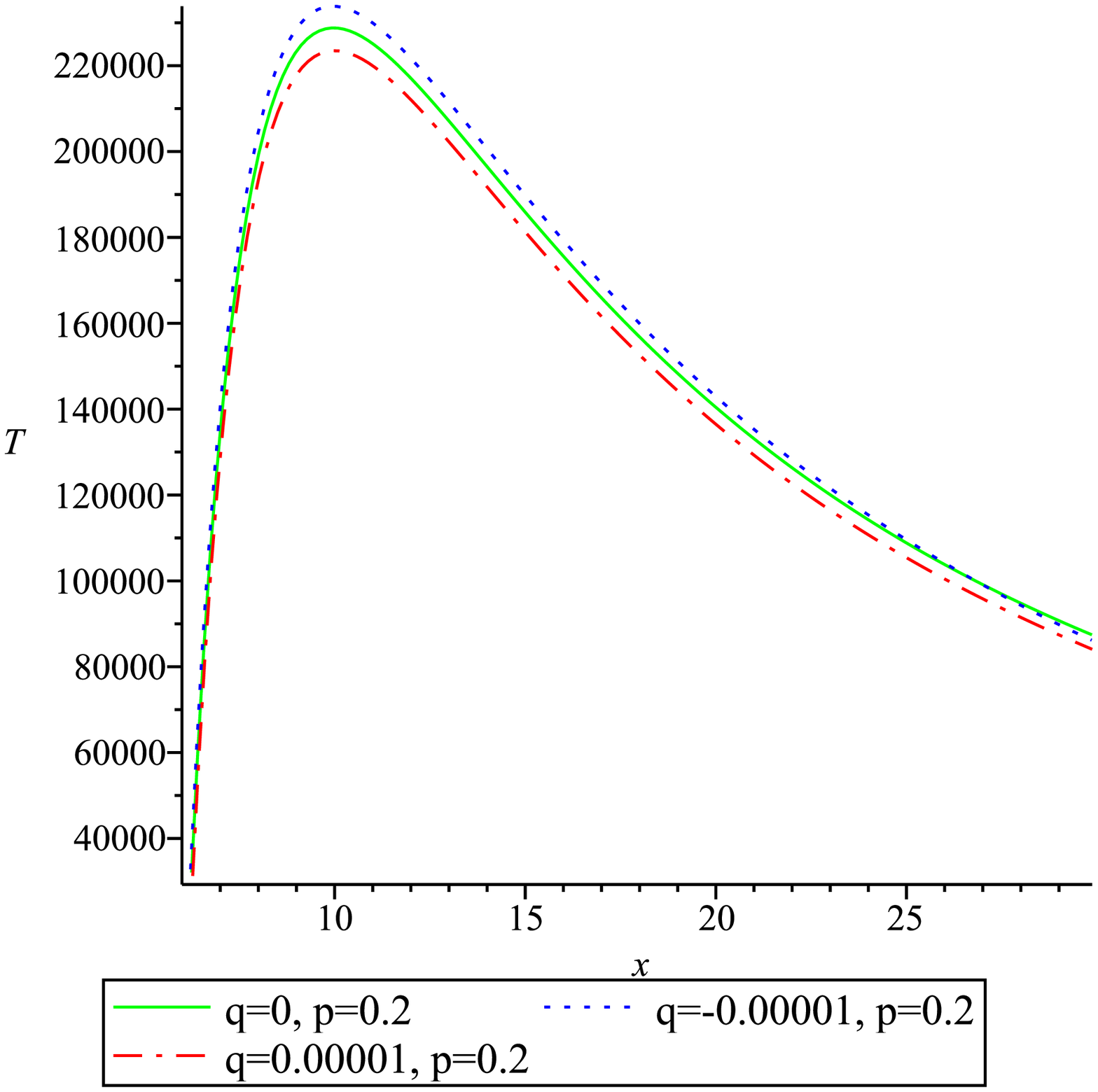}
    \includegraphics[width=0.525\textwidth]{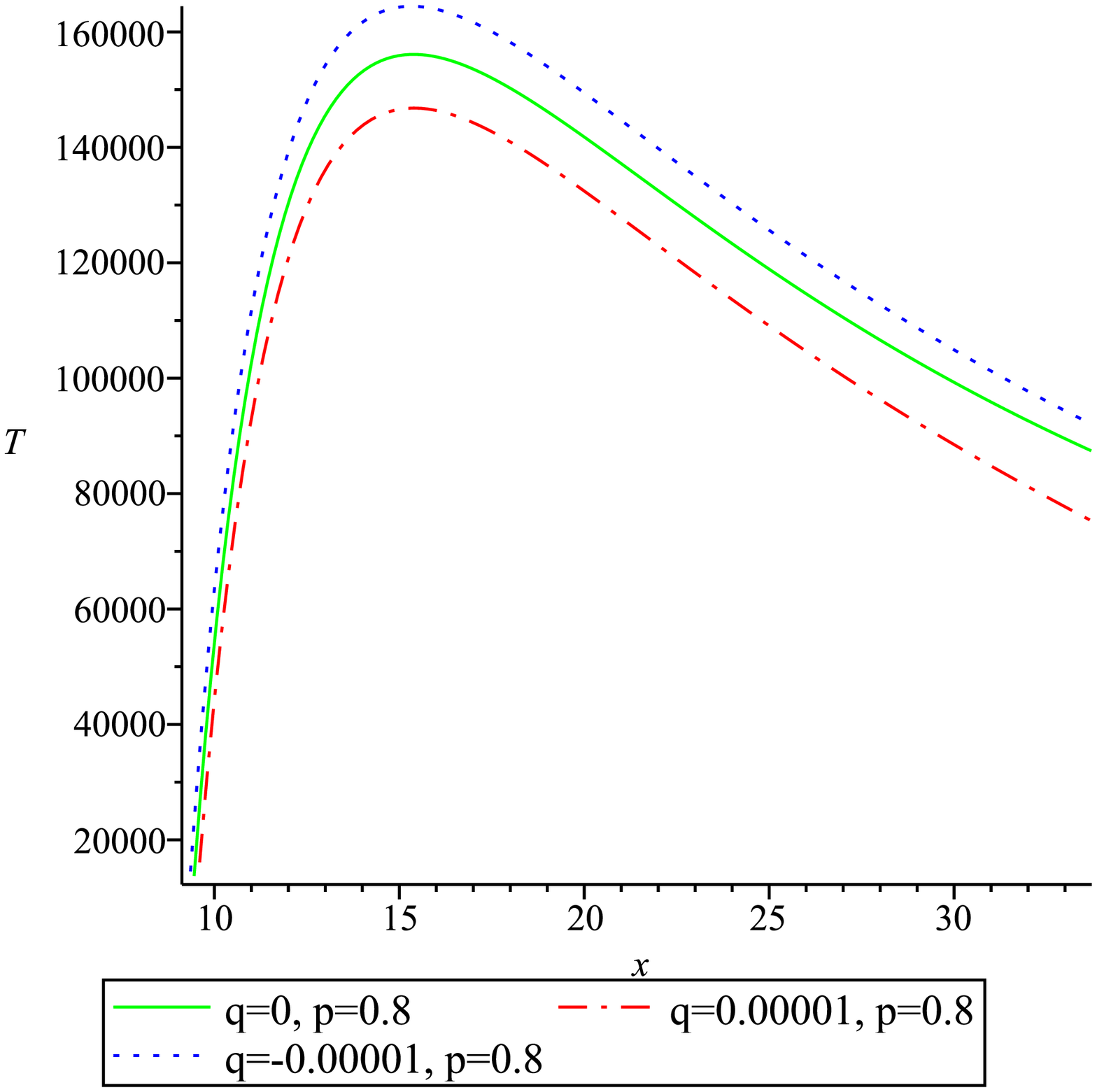}
    \includegraphics[width=0.525\textwidth]{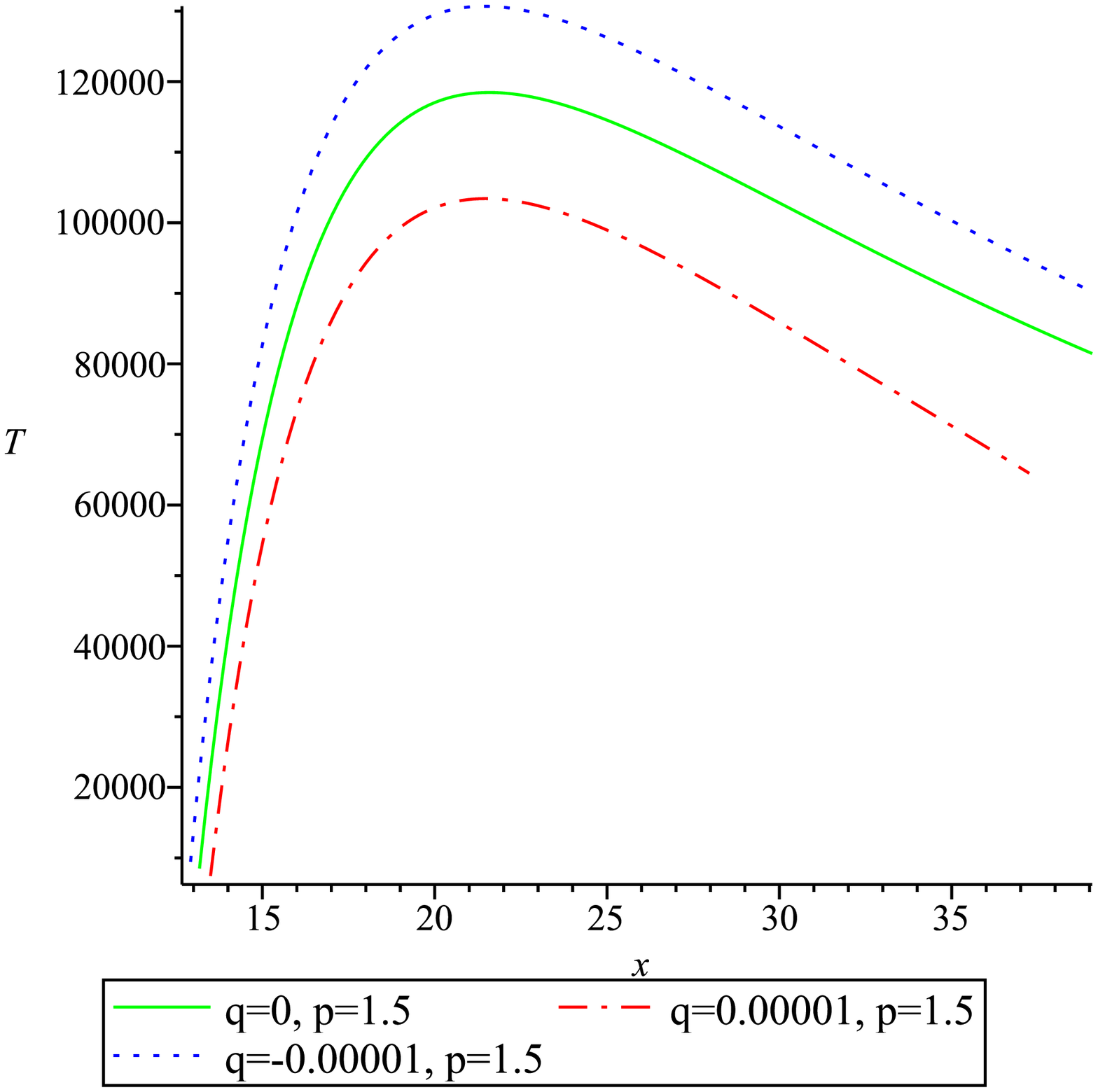}
    \caption{\label{tplot} Temperature $T$ $(K)$ for $\rm p<0$, $\rm p=0$ and $\rm p>0$.}
\end{figure}

\begin{figure}
    \includegraphics[width=0.52\textwidth]{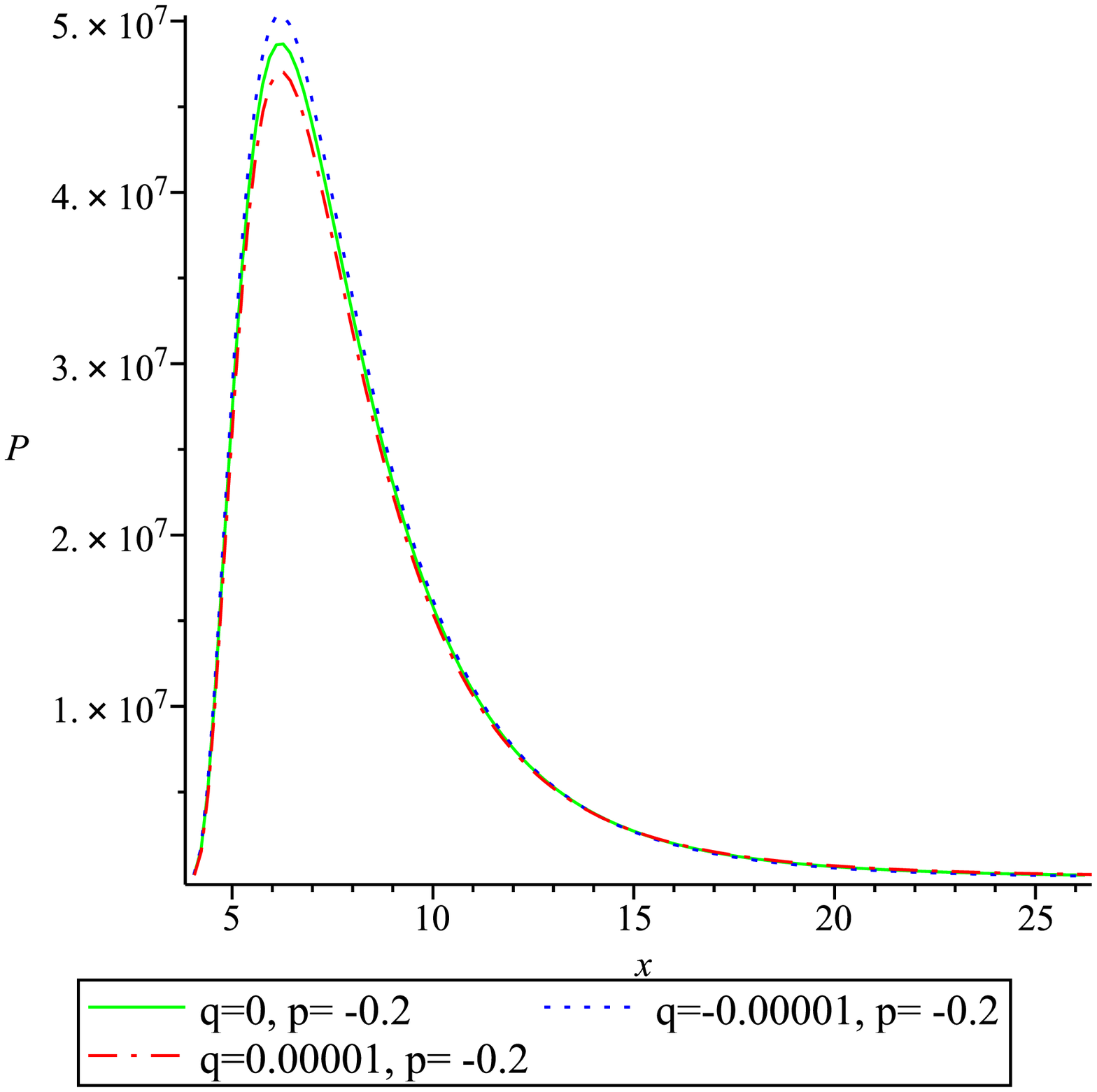}
       \includegraphics[width=0.52\textwidth]{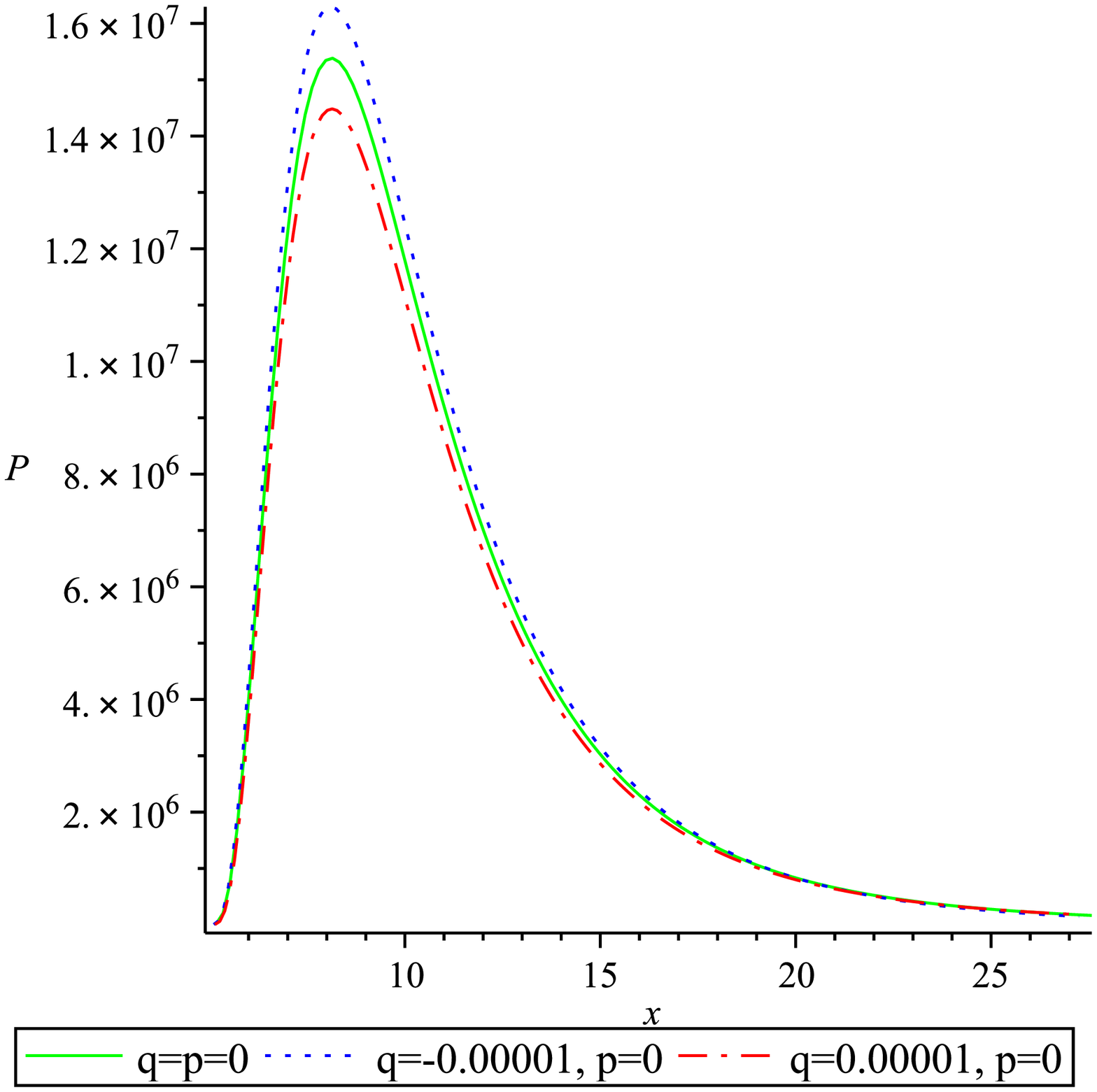}
        \includegraphics[width=0.52\textwidth]{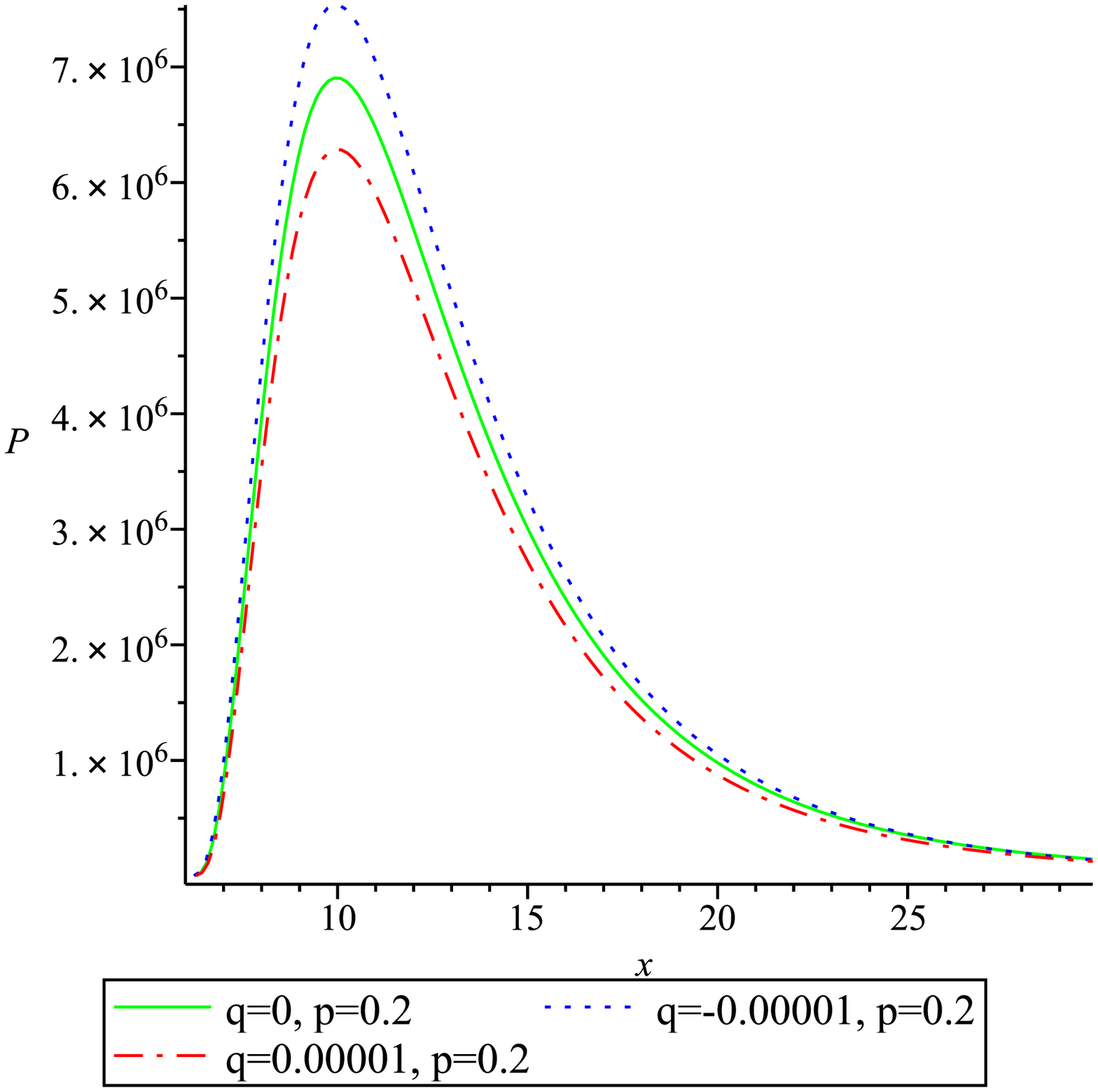}
    \includegraphics[width=0.52\textwidth]{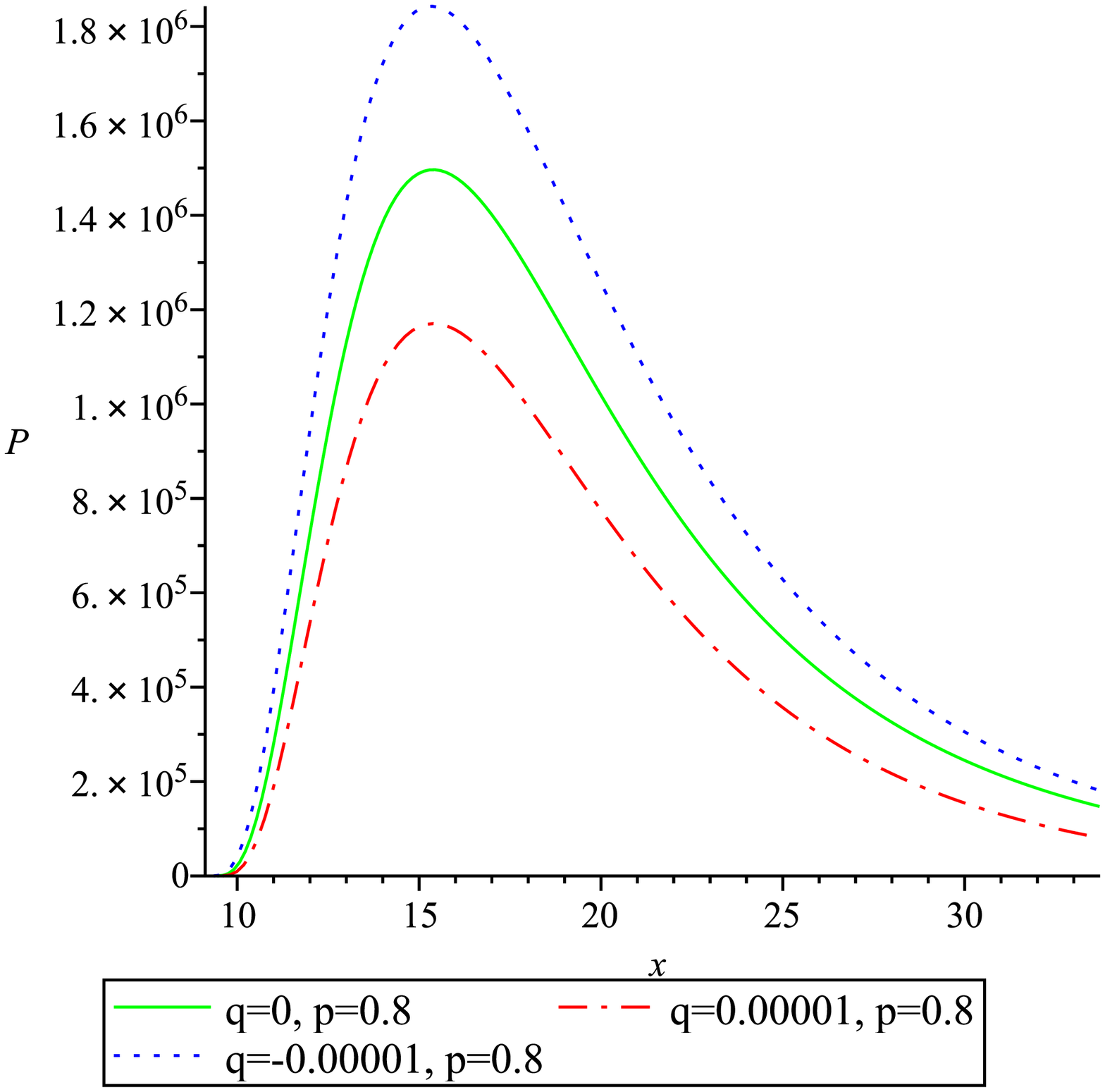}
     \includegraphics[width=0.52\textwidth]{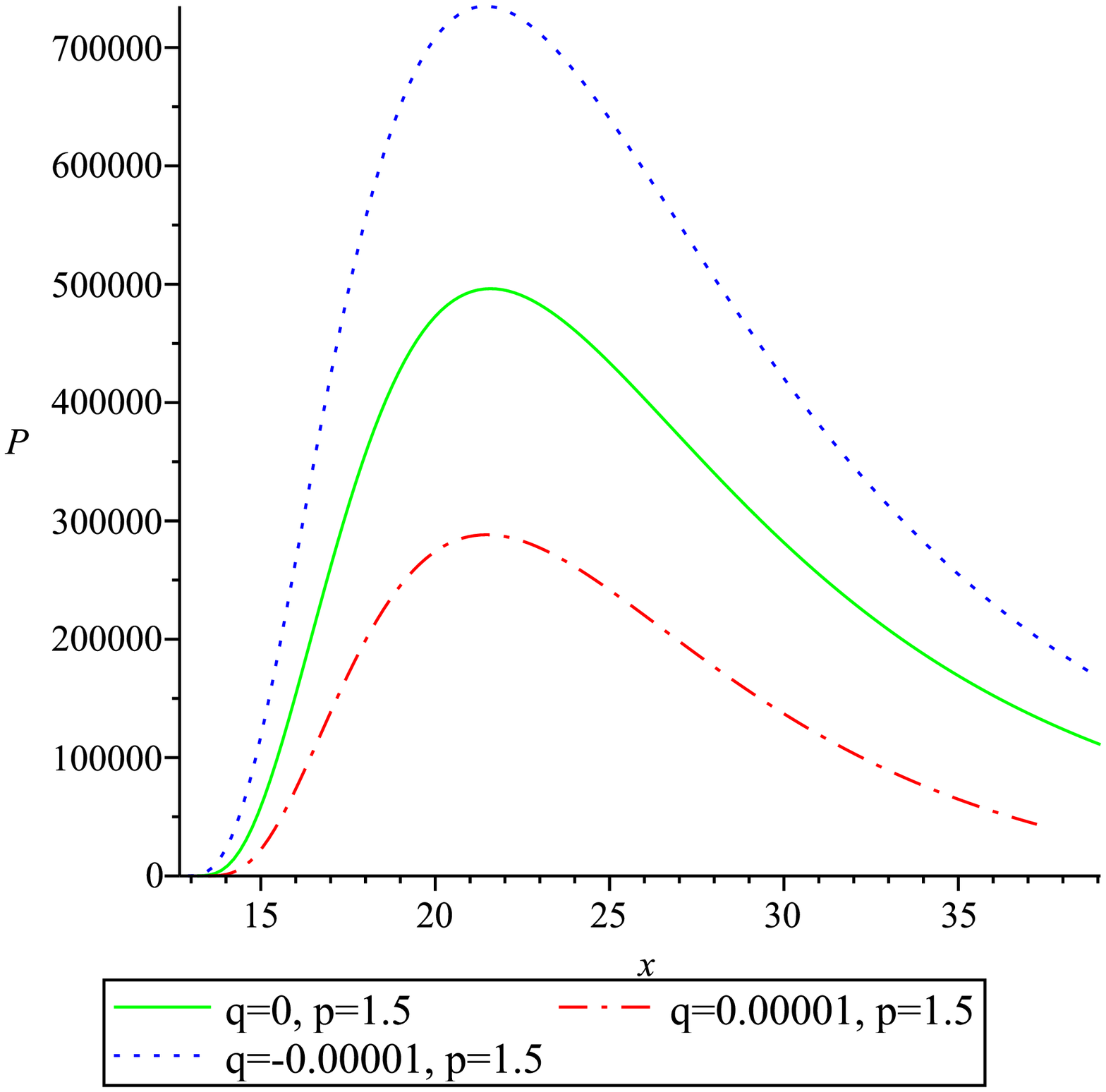}
    \caption{\label{pplot} Pressure $P$ $(\rm g/s^2cm)$, for $\rm p<0$, $\rm p=0$ and $\rm p>0$.}
\end{figure}

\begin{figure}
    \includegraphics[width=0.5\textwidth]{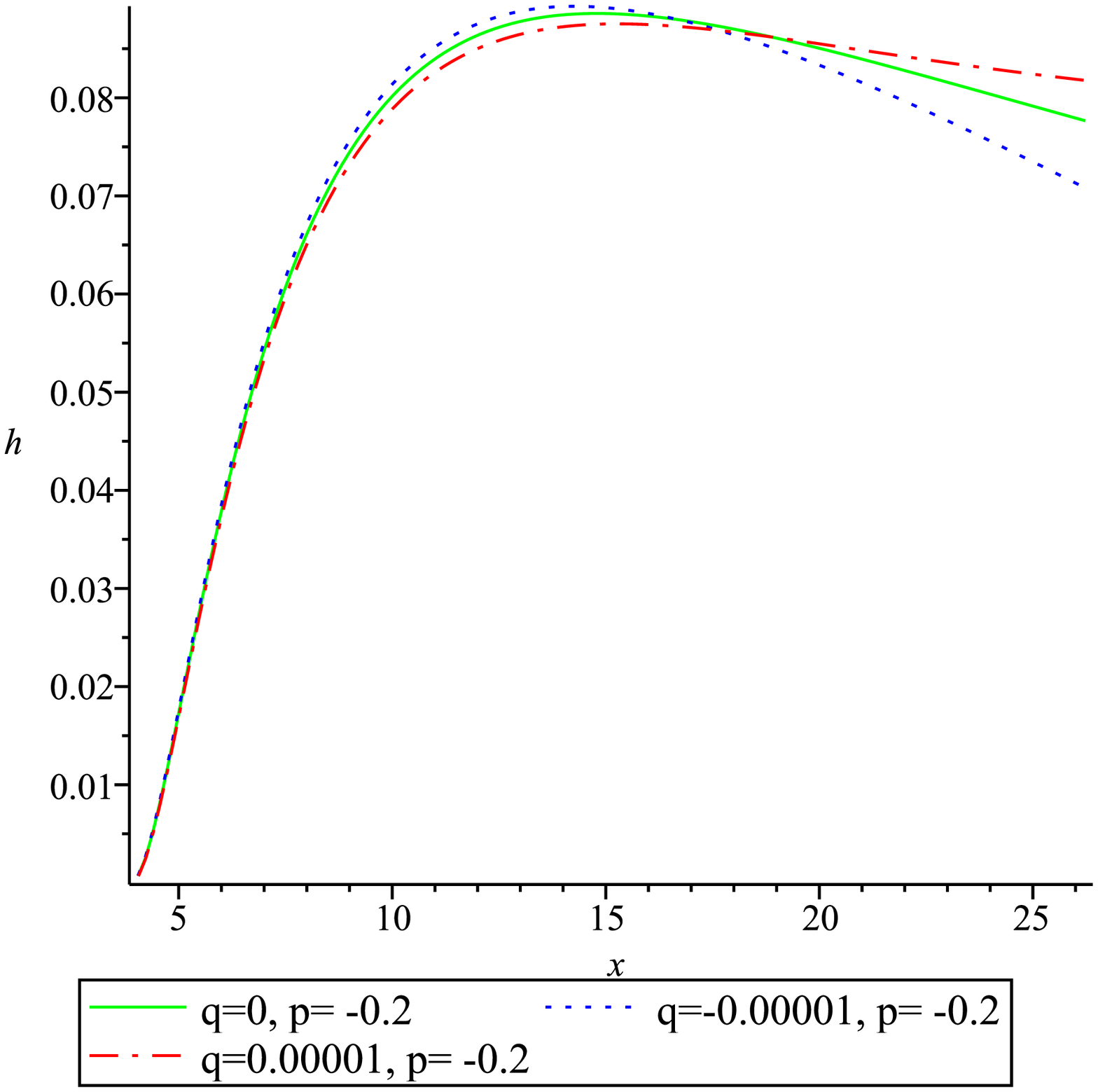}
    \includegraphics[width=0.5\textwidth]{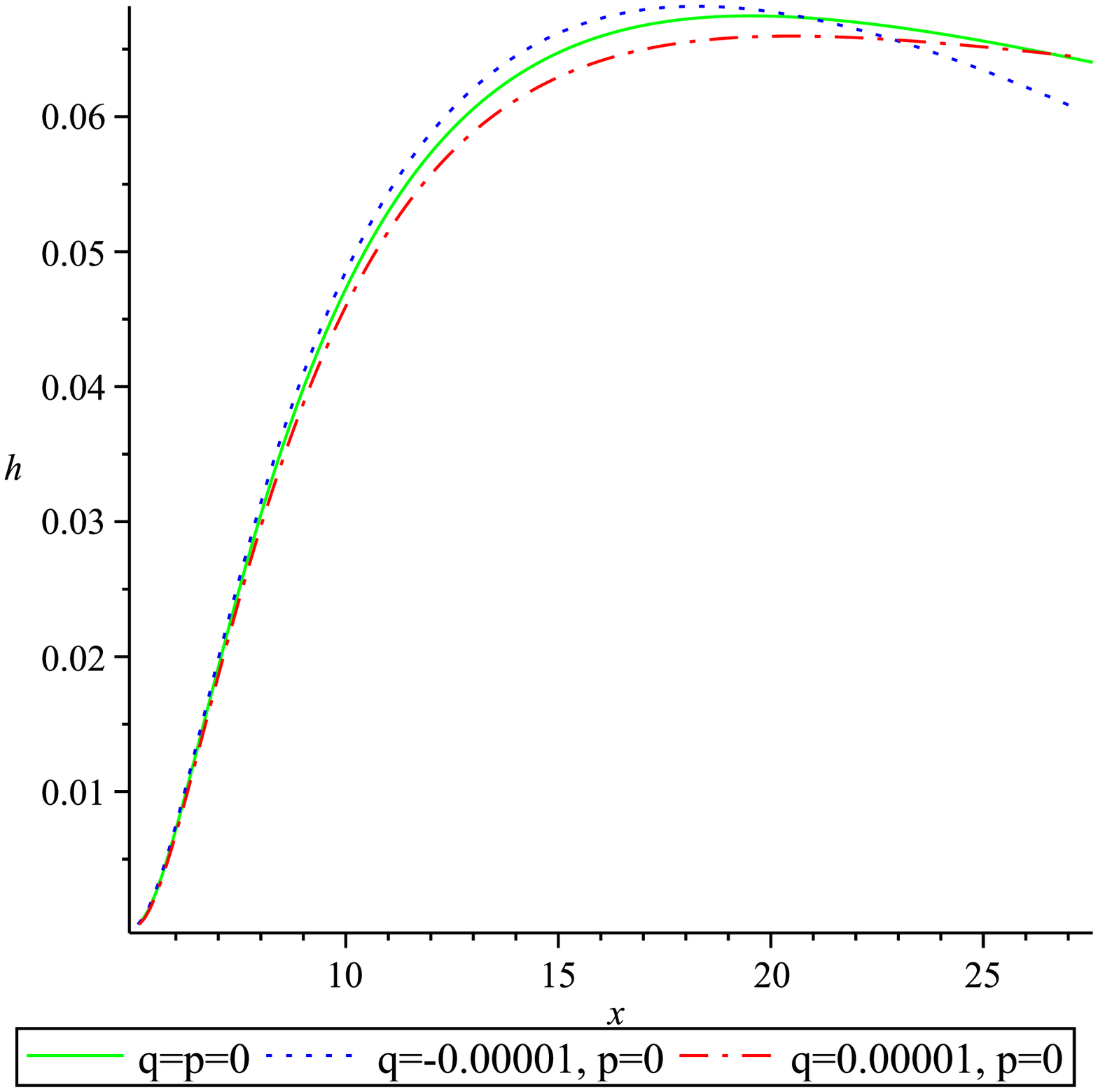}
        \includegraphics[width=0.5\textwidth]{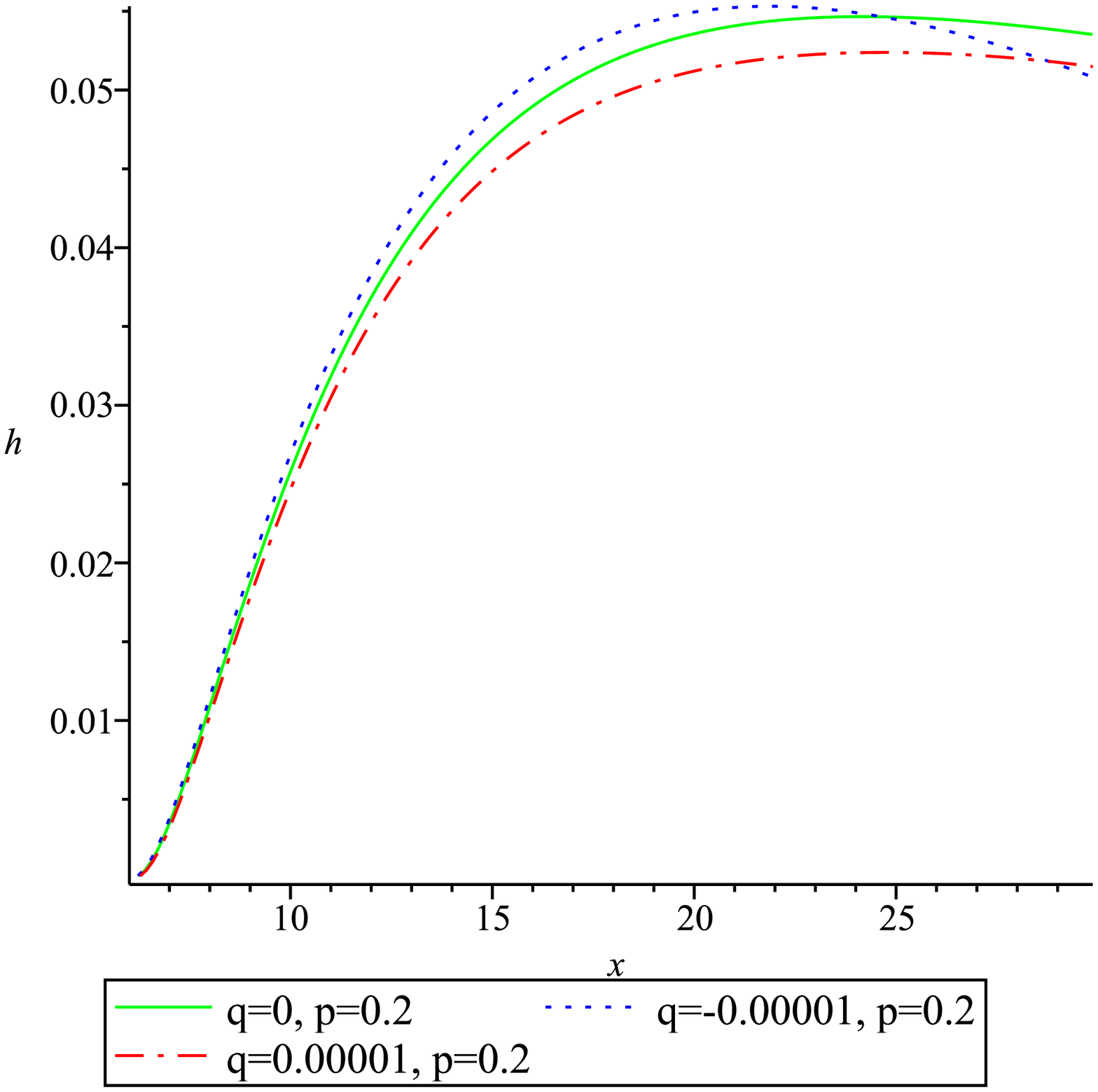}
    \includegraphics[width=0.5\textwidth]{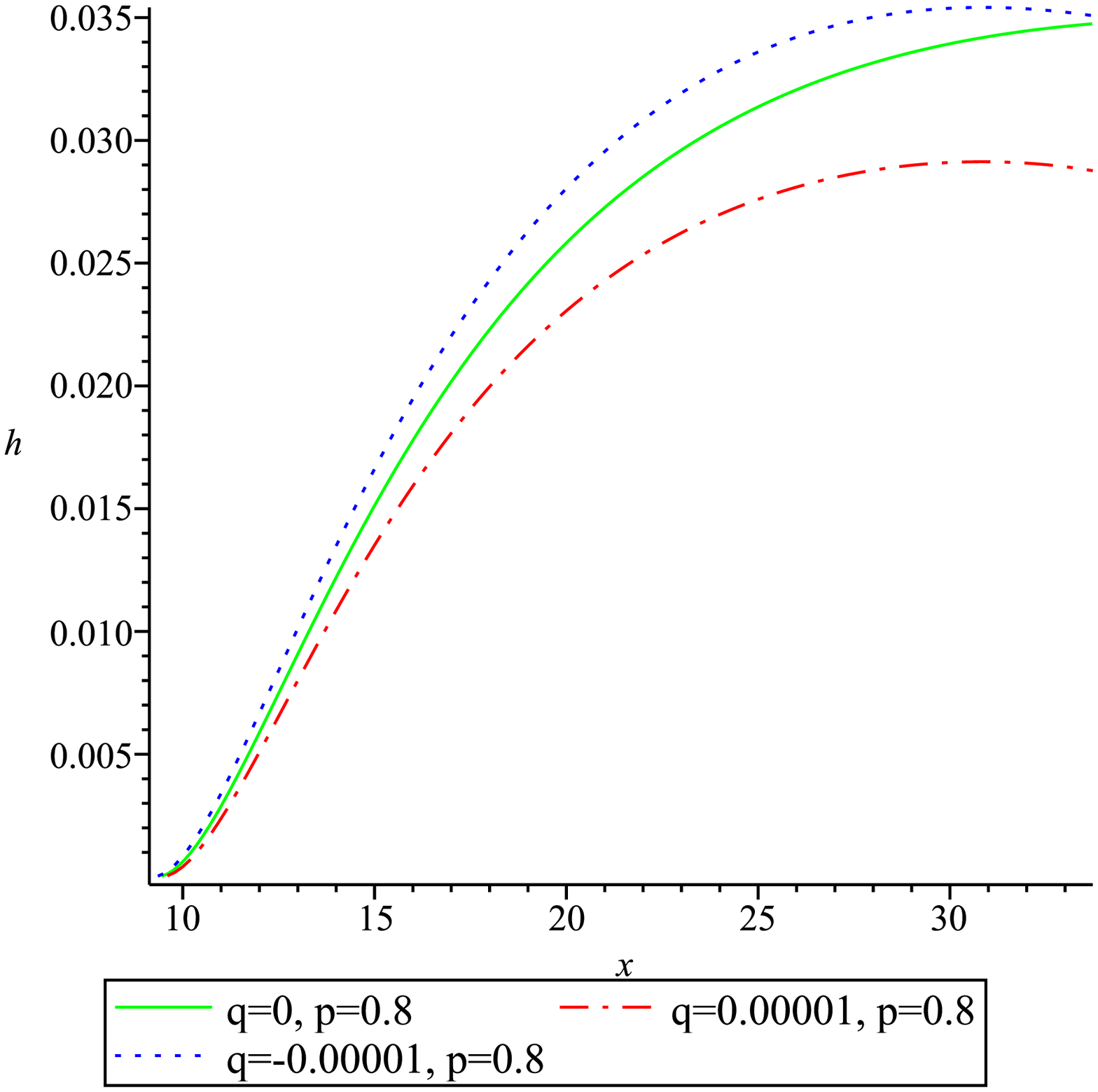}
    \includegraphics[width=0.5\textwidth]{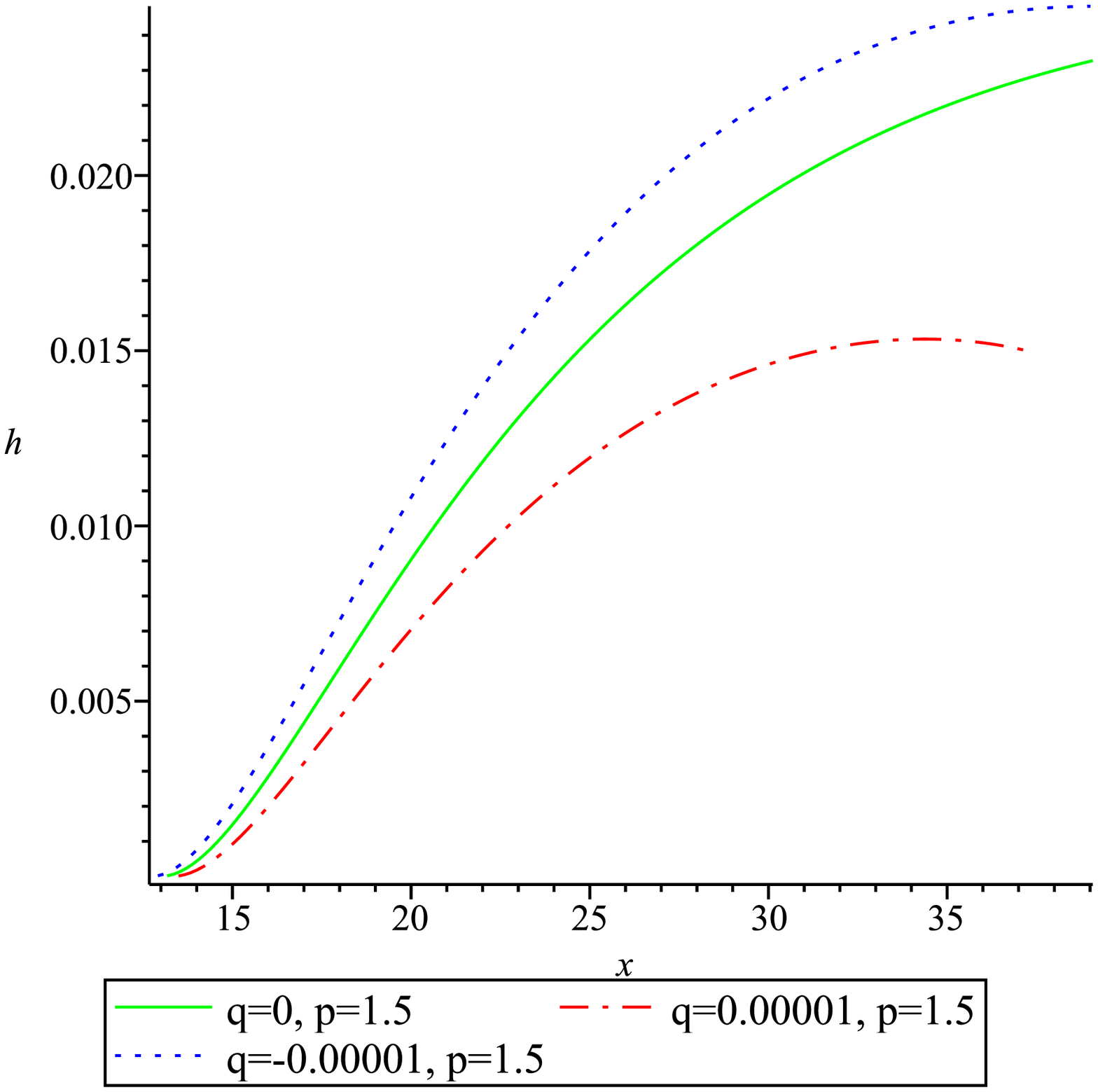}
    \caption{\label{hplot} Height scale of the disc $h$ for $\rm p<0$, $\rm p=0$ and $\rm p>0$.}
\end{figure}

In the distorted case, in order to see the difference clearly, these quantities were plotted separately for Schwarzschild $\rm p=0$, and the chosen negative and positive quadrupoles $\rm p$, in the region that enable us to compare these cases together in Figures \ref{fplot} - \ref{hplot}. In general, the effect of external quadrupole $\rm q$ for a fixed $\rm p$ at small radii, say close to the ISCO, has a rather small impact on the results. The deviations between the case of $(\rm q=0, p)$ and different choices of the quadrupole $\rm q$ become more clear in larger radii. In addition, in all quantities the pick, for a fixed $\rm p$, and negative $\rm q$ is higher than the undistorted one $(\rm q=0, \rm p)$, and for the positive $\rm q$ is lower than the undistorted one. Also, it is clearly seen from the plots that as the quadrupole $\rm p$ become larger the effect of these distortions also become more significant.

The radiation flux $F$, is plotted over distance in Figure \ref{fplot}. Also, the result has a similar pattern to ones for temperature $T$ and the pressure $P$, shown in Figure \ref{tplot} and \ref{pplot}. As it was mentioned earlier, in the case of a positive quadrupole $\rm q$, we see a more rapid descent at larger $x$ compared to the undistorted one $\rm q=0$, for any chosen value of $\rm p$, also for the negative quadrupole moment there is a slower descent with respect to undistorted case. Also at smaller radii, for negative quadrupole $\rm q$ there is a sharper ascent and descent with a higher pick in contrast to the undistorted and positive $\rm q$ case. However, this may not be a surprise, because also in the case of negative quadrupole of the external matter $\rm q$, the inner edge of the disc, ISCO, is getting closer to the horizon even very small since the values for quadrupole $\rm q$ are relatively small (see table \ref{T1}), and we expect to have more intense effects at the inner part. Also, an analysis shows that we have fairly more deviation from the undistorted case for the radiation and pressure, Figure \ref{fplot} and \ref{pplot} rather than temperature Figure \ref{tplot}.
 

Figure \ref{hplot} shows the height scale $h$ of the disc over $x$. In general, the height of the disc gets thicker and after reaching some maximum, gradually becomes thinner. Although this slope changing is less manifest for positive quadrupoles $\rm q$, it is stronger in the case of the negative quadrupole. In this plot, in contrast to others, the shift of pick in distorted cases, for any fixed value of $\rm p$ is more noticeable. In this respect, for a negative quadrupole $\rm q$, the position of the maximum height is shifted to the smaller radii, while the height of the maximum increases. For a positive quadrupole the situation is reversed and the maximum is shifted to the right and its height decreases. 
However, as we have seen in Section \ref{sec:eq}, in the case of negative quadrupole moment $\rm q$, ISCO is closer to the horizon and we expect to see this behaviour.

\begin{figure}
        \includegraphics[width=0.48\textwidth]{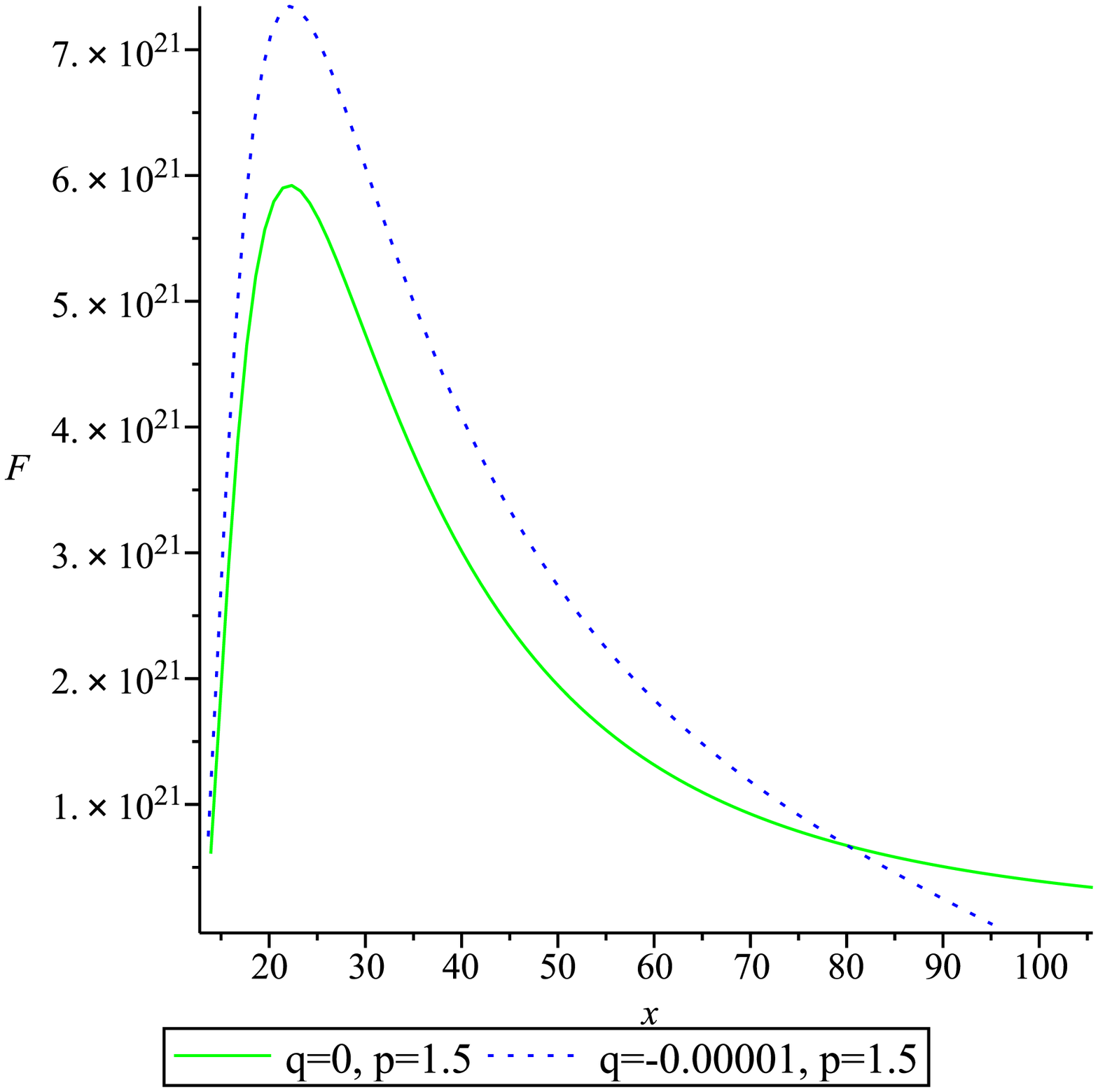}
        \includegraphics[width=0.48\textwidth]{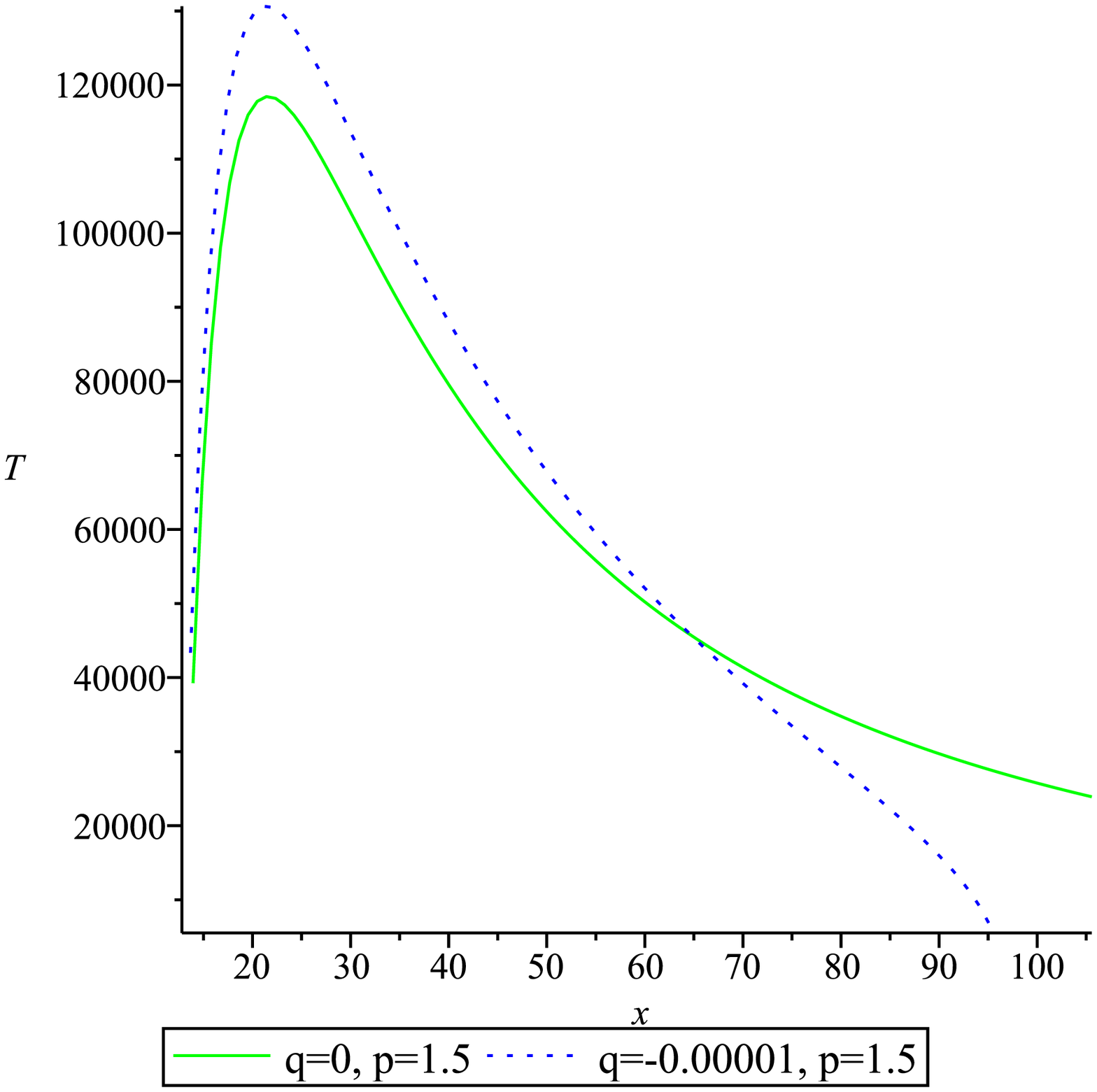}
        \includegraphics[width=0.48\textwidth]{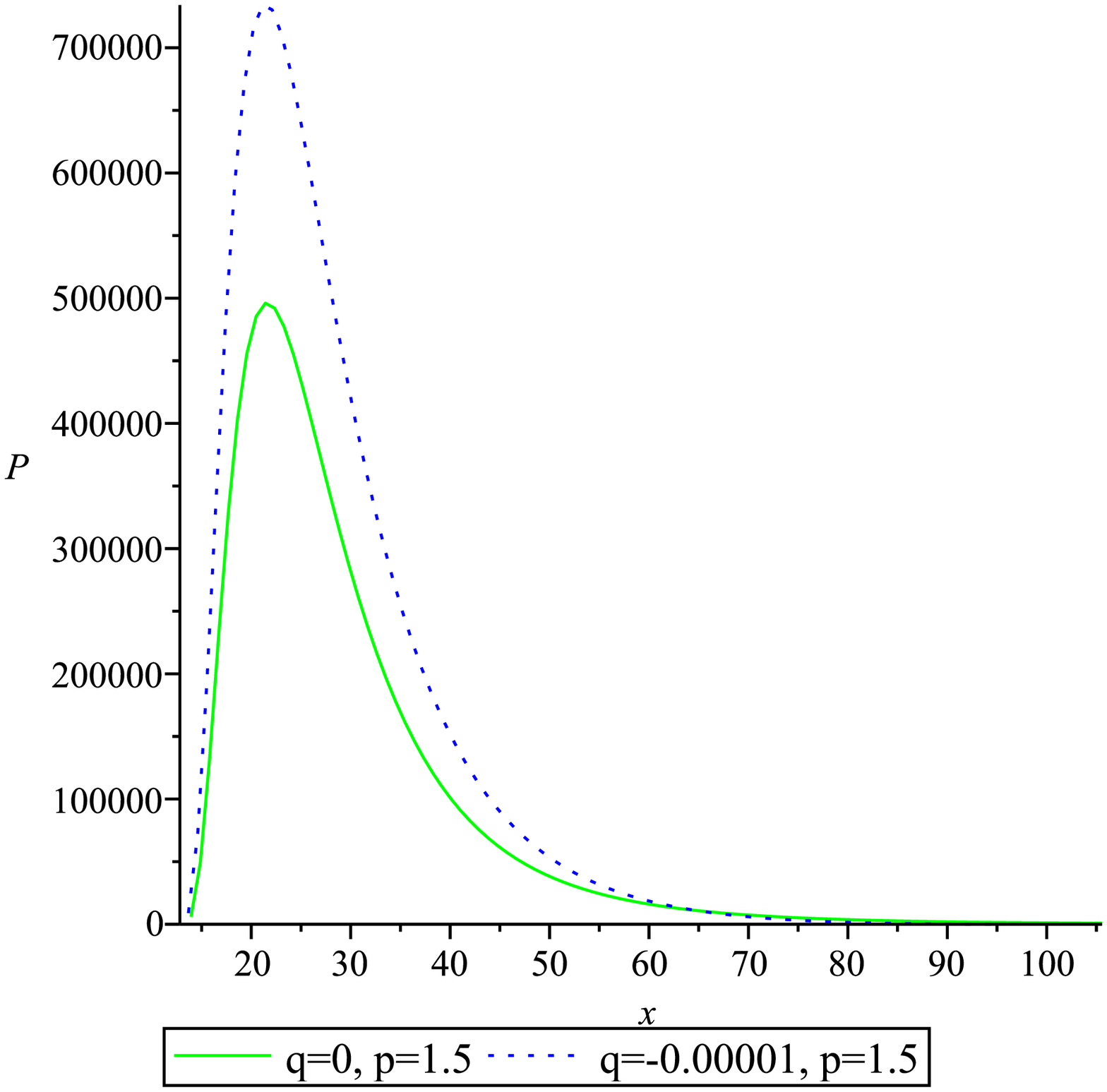}
         \includegraphics[width=0.48\textwidth]{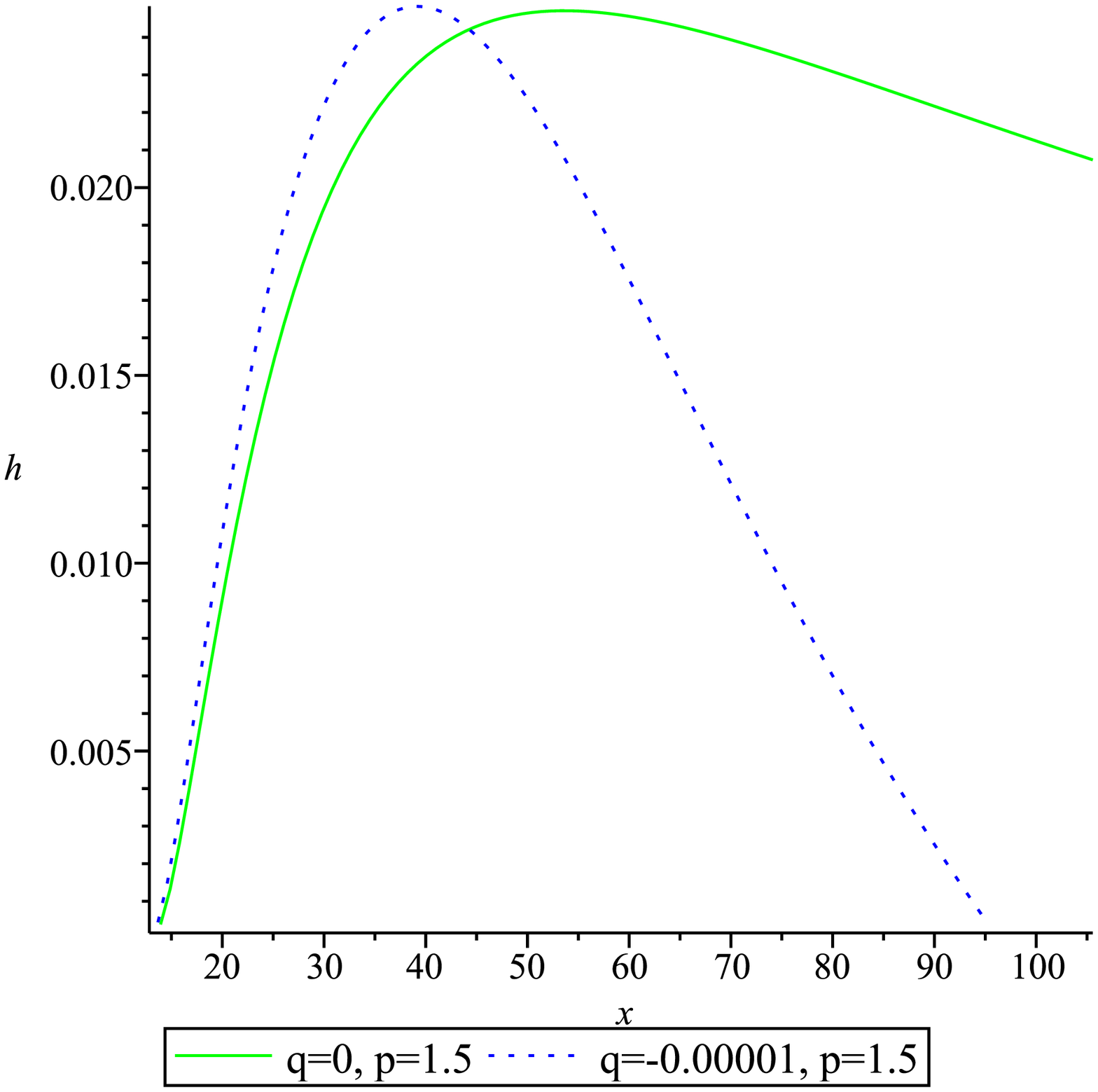}
     \caption{\label{ftphs10}Radiation flux $F$ $(\rm g/s^3)$, temperature $T$ $(K)$, pressure $P$ $(\rm g/s^2cm)$, height scale $h$, for $\rm p=1.5$.}
\end{figure}

In fact, the valid region is more restricted for the positive quadrupole as we have seen from the shear rate analysis in Section \ref{sec:eq}. Thus, in order to see the difference between the case of negative quadrupole $\rm q<0$ and $\rm q=0$, at larger radii these quantities were plotted in Figure \ref{ftphs10}. In this Figure the solid line presents the distorted Schwarzschild case, while the dotted line is plotted distorted naked singularity with deformation parameter $\rm p=1.5$. As the valid range in these cases is comparable, also the quadrupole $\rm p$ is fairly large, we have seen a clear diversion from the static black hole. The more interesting behaviour is for height scale that in this case degrade much faster than the height of a black hole.

\section{Summary and Conclusion}\label{sec:discuss}

In this paper, we have analyzed structure of the thin accretion disc model around distorted and undistorted naked singularities and compared the results with the Schwarzschild black hole via exercising relatively small quadrupoles. In fact, we have shown that the presence of a quadrupole changes drastically the geometric properties of the accretion disc, thus the disc structure in the background of a naked singularity would be significantly different and clearly distinguishable in observation.

Further, in this work the emerging observational perspectives have discussed, with particular reference to the properties of accretion disk. Also, we have derived some physical conditions to determine the upper limit on the valid region depending on quadrupole parameters, as this solution is only valid locally. In general, an analysis shows that the result for a given $ \rm p$, say $\rm p_0$, at small radii close to the ISCO for a nonvanishing external quadrupole $\rm q\neq 0$, has a small effect on the results. However, the deviations between the undistorted naked singularity $(\rm q=0, \rm p_0)$ and different choices of the distortion background, $(\rm q\neq 0, \rm p_0)$ become stronger for larger radii. Also, as the magnitude of the quadrupole $|\rm q|$ approach to zero, we get a wider valid range, as is expected from the $(\rm q=0, \rm p)$ limit for each choice of $\rm p$. Furthermore, the pick of the mentioned quantities for $(\rm q<0, \rm p_0)$ is generally higher than the $(\rm q=0, \rm p_0)$, and the situation is reverse for $(\rm q>0, \rm p_0)$. In addition, the intensity of effects due to distortion background is a monotonically increasing function of $\rm p$. In this respect, a disc around a naked
singularity is much more luminous than one around a regular black hole. Moreover, as $\rm p$ becomes larger the deviation between the distorted and undistorted cases becomes more strong. In this way, utilizing the quadrupole moments can open the possibility of taking them as additional physical degrees of freedom that could be used to link observational phenomena to the models. Especially, as we have some freedom in choosing these parameters, we may decide on ones that minimize computational time and numerical errors in the simulations. 

In summary, the observational characteristic of the extremely strong curvature is in principle distinguishable in many respects with other compact objects in the eyes of accretion disc models. Indeed, this research area associates with our understanding of black hole physics today, its astrophysical applications, and its implication in cosmology and possible quantum theories of gravity. In the next step, one may study the other disc models in this background. Also, the quantities like radiation flux can be analyzed as Gaussian distributions and derive relevant probability functions for them in an stochastically analysis, which may help to understand them better.

\vspace{0.5cm}

\section{Acknowledgements}
This work is initiated in the GR22 conference with a question by Prof. Ramesh Narayan and constructive discussions afterwards. The author gratefully acknowledges him. Also, thanks to the research training group GRK 1620 ”Models of Gravity”, funded by the German Research Foundation (DFG).

\onecolumngrid

\begin{table*}
  \begin{threeparttable}
 \caption{The minimum and the maximum of $\rm q$ and the place of ISCO for different values of $\rm p$.}\label{ta:upper_bound}
  \begin{center}\label{T1}
 \begin{tabular}{|p{1.5cm}||p{3cm}|p{2cm}|p{3cm}|p{2cm}||p{2cm}|} 
 \hline 
$\rm p$&$\rm q_{\rm min}$ &$ x_{\rm q-min}$&${\rm q}_{\rm max}$&$ x_{\rm q-max}$&$x_{\rm {q=0}}$\\
 \hline \hline
   -0.5   & -0.2499996   &  1.000001 & 0.0021728  &  2.58141 & 2.00000 \\\hline
  -0.49   & -0.1757730   & 1.132032 & 0.0019932 & 2.68029   &  2.07818  \\\hline
   -0.4   & -0.0805014   & 1.550384 & 0.0011090 & 3.47165   &  2.69443  \\ \hline
  -0.2   & -0.0355583 & 2.240655 & 0.0004991 & 5.01010   &  3.88324 \\ \hline
  0      & -0.0209443  & 2.879401 & 0.0002927 & 6.45602    &  5.00000  \\ \hline
  0.2    & -0.0139685   & 3.502152 & 0.0001945 & 7.86682   &  6.08998   \\ \hline
  0.5    & -0.0086651   & 4.423403 & 0.0001202 & 9.95281   &  7.70156  \\ \hline
  0.8    & -0.0059211   & 5.336801 & 0.0000820 & 12.02010  &  9.29872  \\ \hline
 9 1      & -0.0047632   & 5.943382 & 0.0000659 & 13.38972  &  10.35890  \\ \hline
  10     & -0.0001533  & 32.985010& 0.0000021 & 74.44744   &  57.57641  \\ \hline
 \hline
 \end{tabular}
 \begin{tablenotes}
    \item 
\end{tablenotes}
\end{center}
\end{threeparttable}
 \end{table*}

\twocolumngrid





\bibliographystyle{unsrt}
\bibliography{bibfilepoledisknaked}

\end{document}